\documentclass[aps,prb,amsmath,twocolumn,amssymb,titlepage,showkeys,superscriptaddress]{revtex4-1}

\usepackage{graphicx}
\usepackage{nicefrac}
\usepackage{amsfonts}

\usepackage{amssymb}
\usepackage{amsmath} 
\usepackage{bbold}

\usepackage{subfigure}
\usepackage{multirow} 
\usepackage{tabularx} 
\usepackage{array}

\usepackage{units}

\usepackage{braket}

\usepackage{bm}
\usepackage{hyperref}

\usepackage{color}
\usepackage[usenames,dvipsnames]{xcolor}

\begin{document}

\title{Interplay between local response and vertex divergences\\ in many-fermion systems with on-site attraction}

\author{D. Springer}
\affiliation{Institute of Solid State Physics, TU Wien,  Austria}
\author{P. Chalupa}
\affiliation{Institute of Solid State Physics,  TU Wien,  Austria}
\author{S. Ciuchi}
\affiliation{Dipartimento di Scienze Fisiche e Chimiche, Universit\`a dell'Aquila, and Istituto dei Sistemi Complessi,
CNR, Coppito-L'Aquila, Italy}
 \author{G. Sangiovanni}
\affiliation{Institut f\"ur Theoretische Physik und Astrophysik and W\"urzburg-Dresden Cluster of Excellence ct.qmat, Universit\"at W\"urzburg, Germany}
 \author{A. Toschi} 
\affiliation{Institute of Solid State Physics, TU Wien, Austria}

\begin{abstract}
We investigate the divergences appearing in the two-particle irreducible vertex functions of many-fermion systems with attractive on-site interactions. By means of dynamical mean-field theory calculations, we determine the location of singularity lines in the phase diagram of the attractive Hubbard model at half-filling, where the local Bethe-Salpeter equations are non invertible.
We find that divergences appear  both in the magnetic and in the density scattering channels. The former affect a sector of suppressed fluctuations and comply with the mapping of the physical susceptibilities of the repulsive case.
At the same time, the appearance of singularities in the density channel of the attractive model demonstrates that vertex divergences can also plague the dominant scattering sectors associated with enhanced local susceptibilities. This constitutes a counterexample to previously proposed interpretations.
Eventually, by exploiting the underlying physical symmetries and a spectral representation of the susceptibilities,
 we clarify the relation between vertex divergences and the local response of the system in different channels.
\end{abstract}
\maketitle


\section{Introduction}
\label{sec:intro}




The rapid development of advanced many-body approaches\cite{Maier2005, Metzner2012,Rohringer2018}, designed to capture the complex physics of correlated fermionic systems, calls for a improved understanding of the  non-relativistic quantum field theory (QFT).
In particular, this applies to the  QFT description of two-particle scattering processes. While these are beyond a standard textbook treatment\cite{Abrikosov1975,Mahan2000},
they represent a crucial ingredient  for (i)  several diagrammatic schemes\cite{Metzner2012,Rohringer2018} built upon approximation of two-particle vertex functions, as well as  for (ii) the description\cite{Liu2012,Toschi2012,Galler2015,Hausoel2017,Kauch2019,Watzenboeck2020} of spectroscopic experiments beyond one-particle photoemission.

The interest in this subject is witnessed by a growing number of studies\cite{Kunes2011,Rohringer2012,Hafermann2014,Gunnarsson2015,Wentzell2016,Kaufmann2017,Tagliavini2018,Thunstroem2018,Reza2019,Krien2019} focusing on the two-particle formalism and the associated algorithmic aspects.
In this work, we make a further step in this direction, by analyzing one surprising property which characterizes the two-particle analog of the self-energy, i.e.\ the irreducible vertex function. We refer here to the occurrence of multiple divergences displayed by this two-particle quantity in the Matsubara frequency domain.
In this respect, we recall that the self-energy expressed as function of Matsubara frequencies diverges only in the ``extreme" case of a Mott-insulating phase, reflecting the complete suppression of the one-particle Green's function. On the contrary, an ubiquitous presence of divergences in the irreducible vertex functions has been recently demonstrated in all fundamental models of  many-electron physics: from the Hubbard atom\cite{Schaefer2016c,Thunstroem2018} to the Falicov-Kimball model\cite{Janis2014,Schaefer2016c}, the Anderson impurity model\cite{Chalupa2018}, the periodic Anderson model\cite{Chalupa2020}, and the Hubbard model\cite{Schaefer2013,Gunnarsson2016,Schaefer2016c}.

These divergences are a manifestation of the breakdown of self-consistent perturbation expansions in QFT and, as it was recently demonstrated, are also directly related\cite{Gunnarsson2017} to the intrinsic multivaluedness\cite{Kozik2015,Stan2015,Tarantino2018} of the Luttinger-Ward functional for interacting many electron systems. 
Mathematically, they correspond to a non-invertibility of the corresponding Bethe-Salpeter equation.  

The physical processes controlling these divergences are not fully clarified yet. 
In fact, they do not appear to be associated to any phase-transition in the systems considered: At low $T$,  they take place well inside of the metallic, Fermi-liquid phases in the AIM\cite{Chalupa2018} and the dynamical mean-field theory solution of the Hubbard model\cite{Schaefer2013,Schaefer2016c}. Heuristically, their occurrence has at first been related\cite{Schaefer2013} to the appearance of kinks in the spectral functions\cite{Byczuk2007} and in the specific heat\cite{Toschi2009,Held2013} or to underlying non-equilibrium properties\cite{Eckstein2010,Schiro2011}.  A recent, more convincing interpretation\cite{Gunnarsson2016,Gunnarsson2017}, however, associates the vertex divergences in a given channel to the suppression of the corresponding physical susceptibility caused  by the electronic interaction. This interpretation works quite satisfactorily in all the cases studied hitherto and can be regarded, to a good extent, as a two-particle generalization of the suppression of the one-particle Green's function by the corresponding self-energy.

In this paper, we study how the divergences of the irreducible vertex functions of the half-filled Hubbard model are transformed by changing the sign of the interaction from $U$ to $-U$. To this end we will perform numerical calculations of two-particle susceptibilities and vertex functions by means of dynamical mean-field theory (DMFT)\cite{Georges1996}. We will interpret our results in terms of the underlying physical symmetries of the model considered and in terms of the mapping between the attractive and repulsive problem. This will allow us to clarify the multifaceted relation between two-particle vertex properties and their divergences and the physical local response of the system. 

These considerations do not only improve our understanding of the physics responsible for the breakdown of the (bold) perturbation expansion\cite{Kozik2015}, but also allow us to make predictions about which kind of vertex divergences can be expected in different physical situations.
Beyond the conceptual progress of an improved mathematical  and physical understanding of the two-particle QFT formalism, our results will be also of particular interest for future developments and applications of several cutting-edge many-electron algorithms (e.g. those based on the parquet formalism\cite{Toschi2007,Ayral2016}, diagrammatic Monte Carlo\cite{Kozik2015}, nested cluster scheme\cite{Vucicevic2018})  beyond the weak-coupling, perturbative regime.

The paper is organized as follows: In Sec.~II, we introduce the basic two-particle formalism needed in our study; in Sec.~III, we present numerical results for the two-particle vertex functions and their divergences in the attractive Hubbard model (IIIA), as well as an interpretation of our findings, based on the mapping of the repulsive case (IIIB) and on a closer inspection of high-$T$ behavior (IIIC); in Sec.~IV we elucidate the relation between vertex divergences and  the physical response of the system by hand of a suitably chosen graphical representation of the generalized two-particle susceptibilities in different channels, and in Sec.~V we discuss possible implications of our results. Our conclusions are summarized in Sec.~VI.

\section{Model and formalism} 
\label{sec:model}
In this work we compute, by means of the dynamical mean-field-theory (DMFT)\cite{Georges1996}, the local two-particle susceptibilities and irreducible vertex functions of both, the attractive and the repulsive Hubbard model, 

\begin{eqnarray}
\label{eq:Hub}
\mathcal{H} = - t \sum_{\langle i, j \rangle,\sigma}  c^{\dagger}_{i,\sigma} c^{\phantom \dagger}_{j,\sigma} + U \sum_{i}  n_{i,\uparrow} n_{i,\downarrow}
\end{eqnarray}
where $c (c^{\dagger})$ are the fermionic annihilation (creation) operators at lattice position $i$  and spin $\sigma$, $t$ is the hopping between next-neighboring sites on a Bethe lattice (with semielliptic DOS of half-bandwidth $D\!=\!2t\!=\!1$),  and the local Hubbard interaction $U$ can take both positive (repulsive interaction) and negative (attractive interaction) values. The chemical potential is kept fixed to $\frac{U}{2}$ to preserve the particle-hole symmetry of the model.

In order to extract irreducible quantities one has to invert the Dyson equation at the one- and the Bethe-Salpeter equation (BSE) as well as the parquet equation at the two-particle level. 

At the one-particle level the self-energy $\Sigma(\nu)$ can be computed from the inversion 
\begin{equation}
\Sigma(\nu) = {\mathcal G}_0^{-1}(\nu) - G^{-1}(\nu), 
\label{eq:Dysoninv}
\end{equation}
of the non-interacting Green's function $ {\mathcal G}_0$ and the interacting impurity Green's function 
$$G(\nu) =  - \int_0^\beta \, d\tau \,  \mbox{e}^{i\nu \tau}  \, \langle T_\tau c(\tau) c^\dagger(0) \rangle$$ 
of the auxiliary AIM associated to the DMFT solution (here $\nu = \pi T (2n+1)$ is a fermionic Matsubara frequency). 
Equation~\eqref{eq:Dysoninv} illustrates that a divergence of $\Sigma(\nu)$ is associated to a complete suppression of $G(\nu)$, which only occurs in the Mott-insulating regime for $T , \nu \rightarrow 0$ .

The analog of $\Sigma$ at the two-particle level\cite{Bickers2004,Rohringer2012} is the irreducible vertex function $\Gamma_{r}$, given in a specific scattering channel $r$ (e.g. density, magnetic, see below). $\Gamma_r$ is obtained by inverting the corresponding BSE
\begin{equation}
\label{eq:BSEinv}
\Gamma_{r}^{\nu \nu'}(\Omega) = \beta^2 \big( [\chi_{r}^{\nu \nu'}(\Omega)]^{-1} - [ \chi_{0}^{\nu \nu'}(\Omega)]^{-1} \big) \ ,
\end{equation}
where the explicit expression of the  generalized susceptibility of the impurity-site in particle-hole notation
\cite{Rohringer2012,Rohringer2018} reads
\begin{eqnarray}
\label{equ:form_gen_chi}
\chi_{\sigma \sigma'}^{\nu \nu'} (\Omega) &=& \int \limits_0^\beta d \tau_1 d\tau_2 d\tau_3 \,
e^{-i\nu \tau_1} 
e^{i(\nu + \Omega)\tau_2} e^{-i(\nu' + \Omega)\tau_3} \nonumber  \\
&\times & [ \langle T_{\tau} c_{\sigma}^{\dagger} (\tau_1)
c_{\sigma}^{\phantom \dagger}(\tau_2) c_{\sigma'}^{\dagger}(\tau_3) c_{\sigma'}^{\phantom \dagger}(0) 
\rangle \\
& -& \langle T_{\tau} c_{\sigma}^{\dagger} (\tau_1)
c_{\sigma}^{\phantom \dagger}(\tau_2) \rangle \langle T_{\tau} c_{\sigma'}^{\dagger}(\tau_3) 
c_{\sigma'}^{\phantom \dagger}(0) \rangle \nonumber] \ .
\end{eqnarray}
Here, $\sigma$ and $\sigma'$ denote the spin directions of the
impurity electrons,  while $\nu$, $\nu'$ and 
$\Omega$ represent two fermionic and one bosonic Matsubara frequency, respectively.  
$\chi_{0}^{\nu \nu'}$ corresponds to the bare bubble given by $-\beta G(\nu) G(\nu + \Omega) \delta_{\nu \nu'}$.
In the case of SU(2) symmetry, the BSE can be diagonalized in the spin sector 
defining the density ($r\!=\!d$) and magnetic ($r\!=\!m$) channel: 
$\chi_{d[m]}^{\nu \nu'} (\Omega) = \chi_{\uparrow \uparrow}^{\nu \nu'} (\Omega) +[-]
\chi_{\uparrow \downarrow}^{\nu \nu' } (\Omega)$.
Similar considerations apply to the particle-particle ($pp$)-sector for which the expression of the generalized susceptibilities
in the corresponding ($pp$) notation can be obtained\cite{Rohringer2012,Rohringer2013a} via a frequency shift of the particle-hole expressions $\chi_{pp,\uparrow\downarrow}^{\nu\nu'}(\Omega) = \chi_{\uparrow\downarrow}^{\nu\nu'}(\Omega -\nu-\nu')$. 
\footnote{With this definitions, a BSE, formally similar to Eq.(\ref{eq:BSEinv}), can be written for the quantity $\chi_{pp}^{\nu\nu'}(\Omega) = \chi_{pp, \uparrow\downarrow}^{\nu (\Omega-\nu')}(\Omega) - \chi_{0,pp}^{\nu\nu'}(\Omega)$ with $ \chi_{0,pp}^{\nu \nu'}= -\beta G(\nu) G(\Omega- \nu') \delta_{\nu \nu'}$.}

The inversion of $\chi_r^{\nu \nu'}$ in Eq.~(\ref{eq:BSEinv}) can also be written in
 terms of its eigenvalue decomposition. For the static case ($\Omega=0$), considered in this work, the 
 explicit expression is 
  \begin{equation}
\label{eq:InvChiEVSpectrum} 
[\chi_{r}^{\nu \nu'}]^{-1} = \sum_\ell V^r_\ell(\nu)^* [\lambda^r_\ell]^{-1} V^r_\ell(\nu') \ ,
\end{equation}
$\lambda^r_\ell$ and $V^r_\ell(\nu)$ being the eigenvalues and the eigenvectors of $\chi_r^{\nu \nu'},$ respectively.
Similar to the one-particle level, where $\Sigma(\nu)\!\rightarrow\!\infty$ directly corresponds to a zero of $G(\nu)$, a divergence of the two-particle self-energy, i.e. the irreducible vertex $\Gamma_r^{\nu \nu'}$, is related to a vanishing eigenvalue $\lambda_\ell^r$ in Eq.~\eqref{eq:InvChiEVSpectrum}. 
Note that this is merely an analogy, since a single vanishing eigenvalue $\lambda_\ell^r$ does not imply a vanishing of the entire $\chi_r^{\nu \nu'}$ matrix. Hence, a divergence of $\Gamma_r$ does not cause the corresponding static ($\Omega=0$) physical susceptibility
\begin{equation}
\chi_{r}= \frac{1}{\beta^2} \sum_{\nu, \nu'}  \, \chi_{r}^{\nu \nu'}(\Omega=0) 
\label{eq:chiphys}
\end{equation}
to vanish as well. 
However, after crossing a divergence line the corresponding eigenvalue $\lambda_\ell^r$ becomes negative, resulting in a negative contribution in the eigenvalue decomposition of the physical susceptibility
\begin{equation}
\label{eq:ChiSpectrum} 
\chi_{r}= \sum_\ell \lambda^r_\ell | \sum_\nu V^r_\ell(\nu) |^2 \ .
\end{equation}
eventually causing a  progressive suppression of the physical fluctuations in the respective channel. In this respect, a divergence of $\Gamma_r^{\nu \nu'}$ followed by the presence of a negative eigenvalue in $\chi_r^{\nu \nu'}$ may be interpreted as  a two-particle analog of the suppression of the single-particle Green's function by the single-particle self-energy \cite{Gunnarsson2016,Gunnarsson2017}.

Indeed, in all previous studies of models with repulsive interactions, negative eigenvalues have exclusively occurred in physical channels that are suppressed upon increasing the interaction strength $U$, namely in the charge and in the particle-particle sectors. 

According to this observation, one may expect that all vertex divergences in models with  attractive interaction will occur in the (suppressed) magnetic channel. This would heuristically be consistent with the known mapping of the physical degrees of freedom (D.o.F.) of the half-filled Hubbard model.
Due to the intrinsic  $O(4) = SU(2) \times SU(2)$ symmetry, the partial particle-hole, or Shiba, transformation\cite{Shiba1972,Micnas1990} 
\begin{equation}
c_{i \uparrow} \rightarrow  c_{i \uparrow} \ \ \text{and} \ \ c_{i\downarrow} \rightarrow  (-1)^{i} c_{i \downarrow}^\dagger \label{eq:shiba}
\end{equation}
acts as a mapping of all physical observables between $U<0$ and $U>0$. 
In particular, the two SU(2) spin ($\vec{S}$) and pseudospin  ($\vec{S}_p$) sectors, which are related to the respective suppressed channels on the attractive and repulsive side, are transformed into each other
\begin{eqnarray}
\mathcal{S}_x &=& \frac{1}{2}[c_\uparrow^\dagger c_\downarrow + c_\downarrow^\dagger c_\uparrow] \leftrightarrow -\frac{1}{2}[c_\uparrow^\dagger c_\downarrow^\dagger + c_\downarrow c_\uparrow] = \mathcal{S}_{p,x} \nonumber \\
\mathcal{S}_y &=& \frac{i}{2}[c_\uparrow^\dagger c_\downarrow - c_\downarrow^\dagger c_\uparrow] \leftrightarrow \ \ \frac{i}{2}[c_\uparrow^\dagger c_\downarrow^\dagger - c_\downarrow c_\uparrow] = \mathcal{S}_{p,y} \label{Eq:SpinMap} \\
\mathcal{S}_z &=& \frac{1}{2}[c_\uparrow^\dagger c_\uparrow - c_\downarrow^\dagger c_\downarrow] \leftrightarrow \frac{1}{2}[c_\uparrow^\dagger c_\uparrow + c_\downarrow^\dagger c_\downarrow - 1] = \mathcal{S}_{p,z} \ . \nonumber 
\end{eqnarray}
These relations between the physical D.o.F. suggest that an analogous mapping may as well apply to the vertex-divergences.
However, as already noted in Refs.~[\onlinecite{Rohringer2012,DelRe2019}], the mapping of  generalized two-particle quantities, and especially of dynamical irreducible vertices, is more complex than Eq.~(\ref{Eq:SpinMap}) would suggest.

We will examine in the next section, first numerically by means of DMFT calculations, and then analytically by means of fundamental symmetry considerations, how this is reflected in the actual occurrence of vertex divergences in the attractive Hubbard model.

\begin{figure*}[t]
\centering\includegraphics[width=0.85\textwidth]{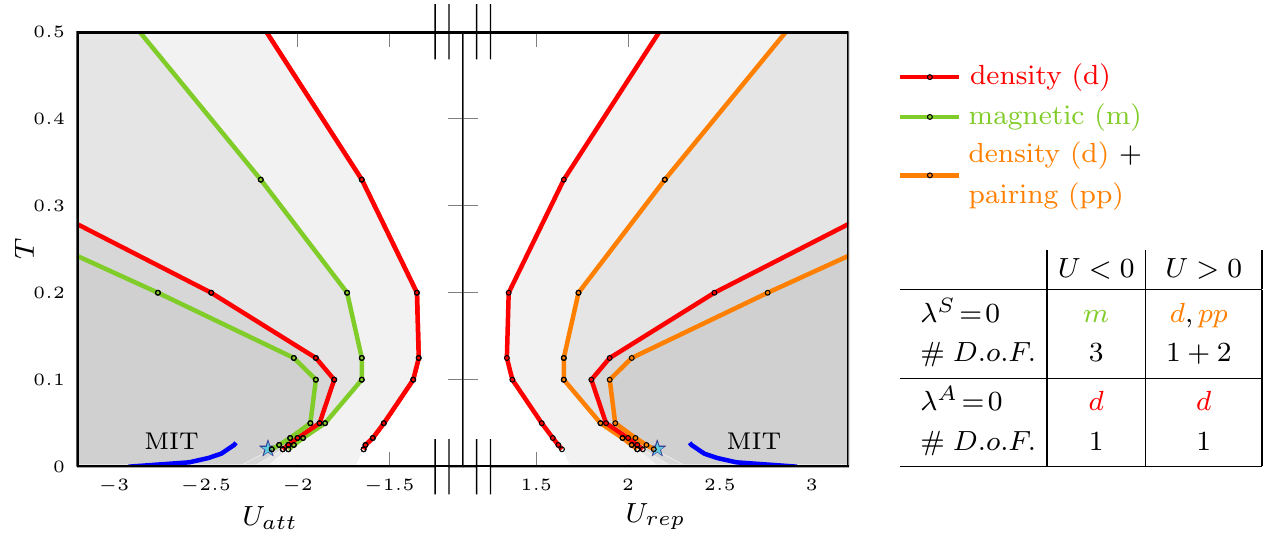}
\caption{Left: Location of the divergences of the irreducible vertex in the different channels along the whole phase-diagram of the attractive {\sl and}  the repulsive half-filled Hubbard model, computed in DMFT.  The comparison of the negative and positive $U$ sectors yields perfectly mirrored divergences in the density channel (red). The simultaneous divergences in the density and particle-particle channel (orange) in the repulsive model are mapped into divergences of the magnetic channel (green) on the attractive side. Stars refers to the position in the phase diagram of the data shown in Fig.~\ref{Fig:Densities}. Right: Schematic sketch  comparing the mapping of the singular eigenvalues $\lambda^S$ [$\lambda^A$]  associated  to symmetric [antisymmetric] eigenvectors to the mapping of  different physical  degrees of freedoms (D.o.F).}
\label{Fig:1}
\end{figure*}


\section{Vertex divergences of the attractive Hubbard model}
\label{sec:phasediag}

\subsection{DMFT results}
\label{subsec:DMFTResults}
We begin our analysis of the vertex functions and their divergences in the attractive Hubbard model by presenting our DMFT calculations at the two-particle level\footnote{For previous DMFT studies of the  attractive Hubbard model at the  one-particle level see, e.g., 
[\onlinecite{Keller2001,Toschi2005a,Toschi2005b,Garg2005,Privitera2010,Koga2011,Tagliavini2016}].}  performed with a continuous time quantum Monte Carlo (CTQMC) impurity solver in the hybridization expansion\cite{Gull2011a}, using the \textit{w}$2$\textit{dynamics}-package \cite{w2dynamics}. 

The main outcome of our DMFT calculations is summarized in the phase-diagram of  Fig.~\ref{Fig:1}, where we report the location of the divergences of  the irreducible vertex $\Gamma_{r}^{\nu \nu'}(\Omega\!=\!0) $ found for different values of the local attraction $U\!<\!0$ and the temperature $T$ (left side), compared  against the corresponding results for the repulsive case $U\!>\!0$ (right side). 
In the large $|U|$ regime our numerical results are consistent with analytical calculations\cite{Thunstroem2018} in the atomic limit. 

Furthermore, in the entire repulsive sector,  we reproduce \footnote{The small differences arise from the different lattice  (here: Bethe lattice) used in the DMFT calculations} the  outcome of previous DMFT studies\cite{Schaefer2013,Schaefer2016c}, finding multiple lines in the $U$-$T$ plane, where the irreducible vertex diverges.
As already observed\cite{Schaefer2016c}, the first divergences are located at moderate repulsion values, well before the Mott-Hubbard MIT. 
With increasing interaction the occurrence of divergence lines becomes more dense. The lines occur in alternating order starting with a divergence in the density channel (red lines) followed by a simultaneous divergence in the density and $pp$ channel (orange lines). 

In the case of attractive interaction we find vertex divergences in the density channel ( red lines), which are perfectly mirrored with respect to the repulsive side. These occur in alternating order with lines of divergences in the magnetic channel (green lines), which mirror the orange divergence lines of the repulsive model. As a consequence, the location of the vertex divergences is highly symmetric when comparing the repulsive and the attractive sides of the phase diagram. 

At first sight this symmetry may appear rather surprising, because the physical properties of a given scattering channel in the repulsive and the attractive model are very different\cite{Micnas1990,Taranto2012,Tagliavini2016}, as dictated by the mapping of the physical degrees of freedom (cf.\ Eq.~\eqref{Eq:SpinMap} and Fig.~\ref{Fig:1}). At a closer look, we can distinguish the three-fold degenerate  divergences found at the orange and green lines, respectively, from the single degenerate divergences found at the red lines, occurring in the density sector only. Specifically, the mapping of the combined divergences in the $pp$ and density sector (orange lines) into divergences of the magnetic sector (green lines) is fully matching our physical expectations: (i) divergences play a role in the suppression of a scattering channel and (ii) they are mapped consistently with the physical D.o.F., i.e. according to Eq.~\eqref{Eq:SpinMap}. At the same time, the perfect mirroring of the  density divergence lines (red) under the $U \!\leftrightarrow \!- U$ transformation looks puzzling, because  (i) for $U \!<\! 0$, these divergences affect a scattering channel associated to a physical susceptibility, which is not suppressed but enhanced by the attractive interaction, and (ii) the physical degrees of freedom associated to the density channel at $U>0$ is mapped onto one of the three spin-components at $U<0$. 

A first understanding of these results is provided by the analysis of the symmetry of the eigenvectors associated to a vanishing eigenvalue ($\lambda^r_{\alpha} =0 ~ \text{for} ~ \ell=\alpha$ in Eq.~(\ref{eq:InvChiEVSpectrum})). In Figure \ref{Fig:2} we compare the shape of eigenvectors following the first and second divergence lines at different temperatures for $U\lessgtr0$. Evidently, the perfect mirroring of divergence lines is also reflected in identical shapes of the corresponding eigenvectors. The singular eigenvectors associated to  all divergences in the density sector only (red lines), display an antisymmetric frequency structure [$V_\ell(-\nu) \! = \! - V_\ell(\nu)$]. In contrast, all other divergence lines (green and orange lines) are associated to frequency symmetric singular eigenvectors [$V_\ell(-\nu) \! = \! V_\ell(\nu)$].

The symmetry of eigenvectors is essential in the calculation of the physical susceptibility, as can be seen in Eq.\,\eqref{eq:ChiSpectrum}. Due to the summation over Matsubara frequencies, the value of $\chi_r$ is independent of any antisymmetric eigenvector, irrespective of whether associated to a positive or a negative eigenvalue. Hence, the appearance of negative eigenvalues in a channel is not necessarily associated to a suppression of the respective physical susceptibility.
While in the repulsive model the occurrence of divergences and the suppression of the respective channel coincide maybe incidentally, our calculations of the  attractive model provide a clear-cut counter-example: the crossing of several divergence lines in the density sector is accompanied by an enhanced susceptibility.

The mirrored location of the vertex divergences, found by our DMFT calculations and, in particular,  their non-trivial relation with the mapping of the physical D.o.F. (s.~Eqs.~\ref{Eq:SpinMap}, at the end of Sec.~II) call for an extension of the attractive-repulsive mapping at the level of the generalized two-particle quantities. This will be presented in the next section and, as we will see, it will play a key role for the interpretation of our numerical results.

\subsection{The role of the underlying symmetries}
\label{subsec:symmetries}

As mentioned at the end of Sec.\,\ref{sec:model}, the mapping of the generalized two-particle quantities is less obvious than the mapping of the physical D.o.F.. 
\begin{figure}[t!]
\includegraphics[width=0.23\textwidth]{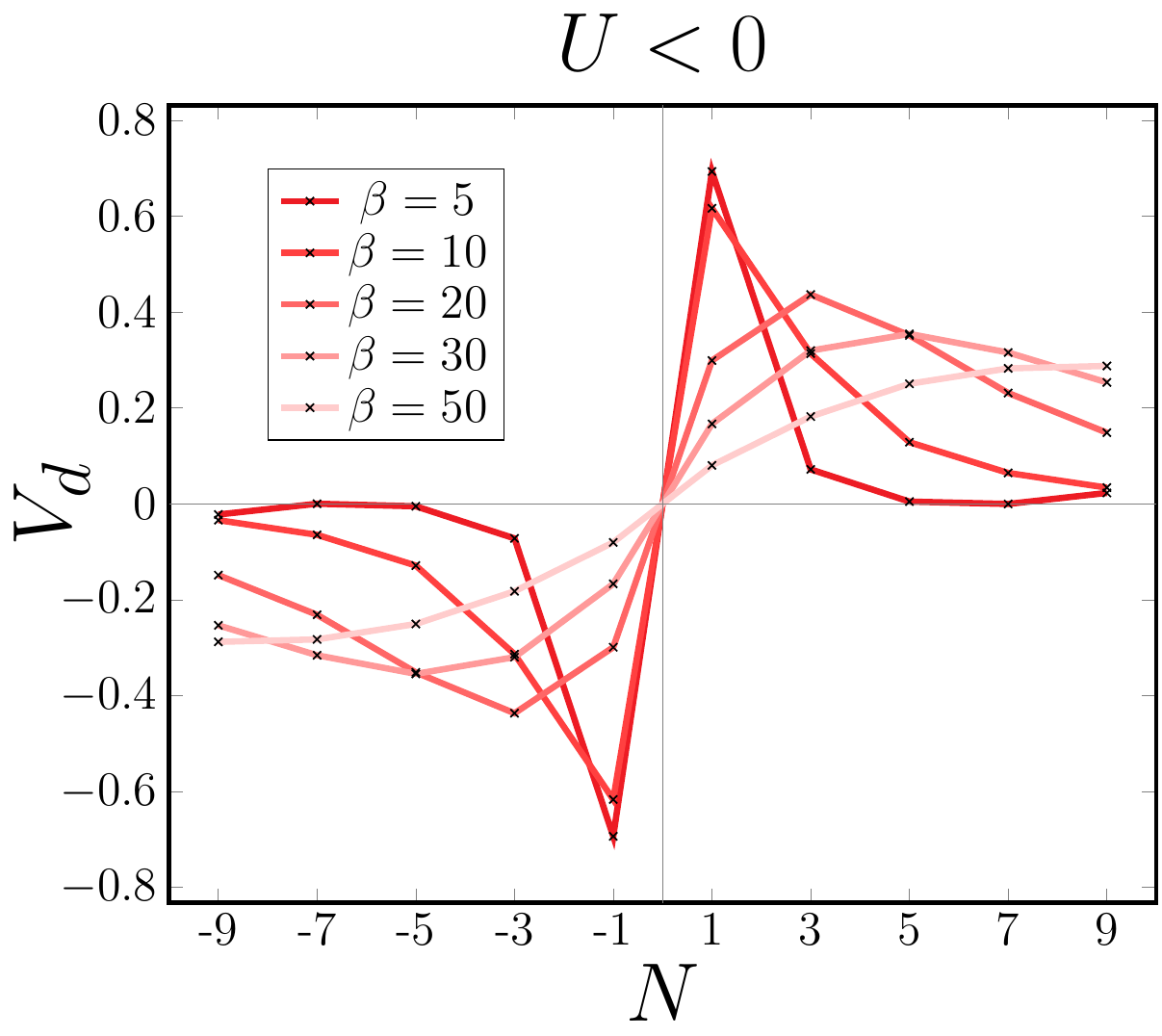}
\includegraphics[width=0.23\textwidth]{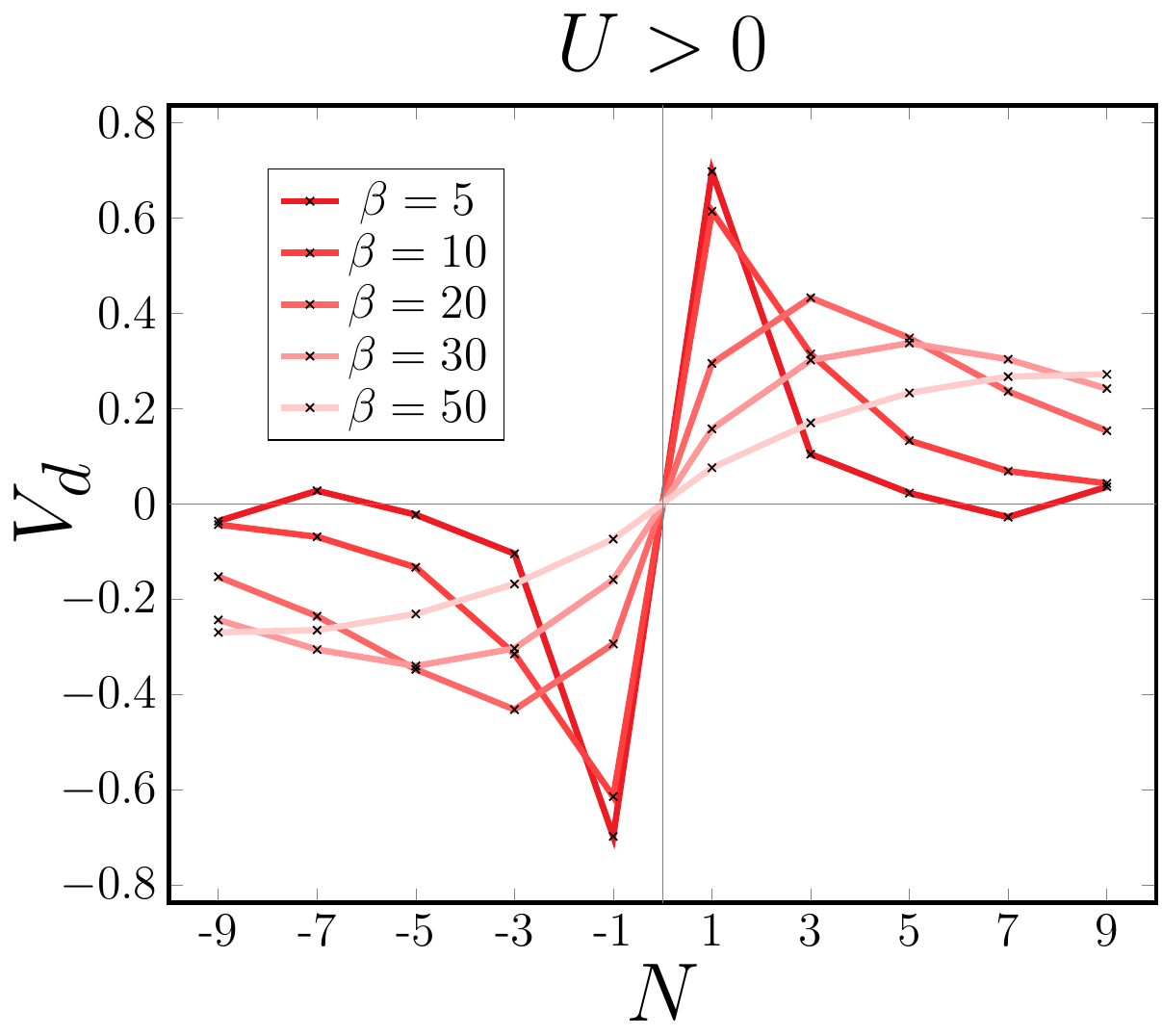}
\\ \vspace{1mm} 
\includegraphics[width=0.23\textwidth]{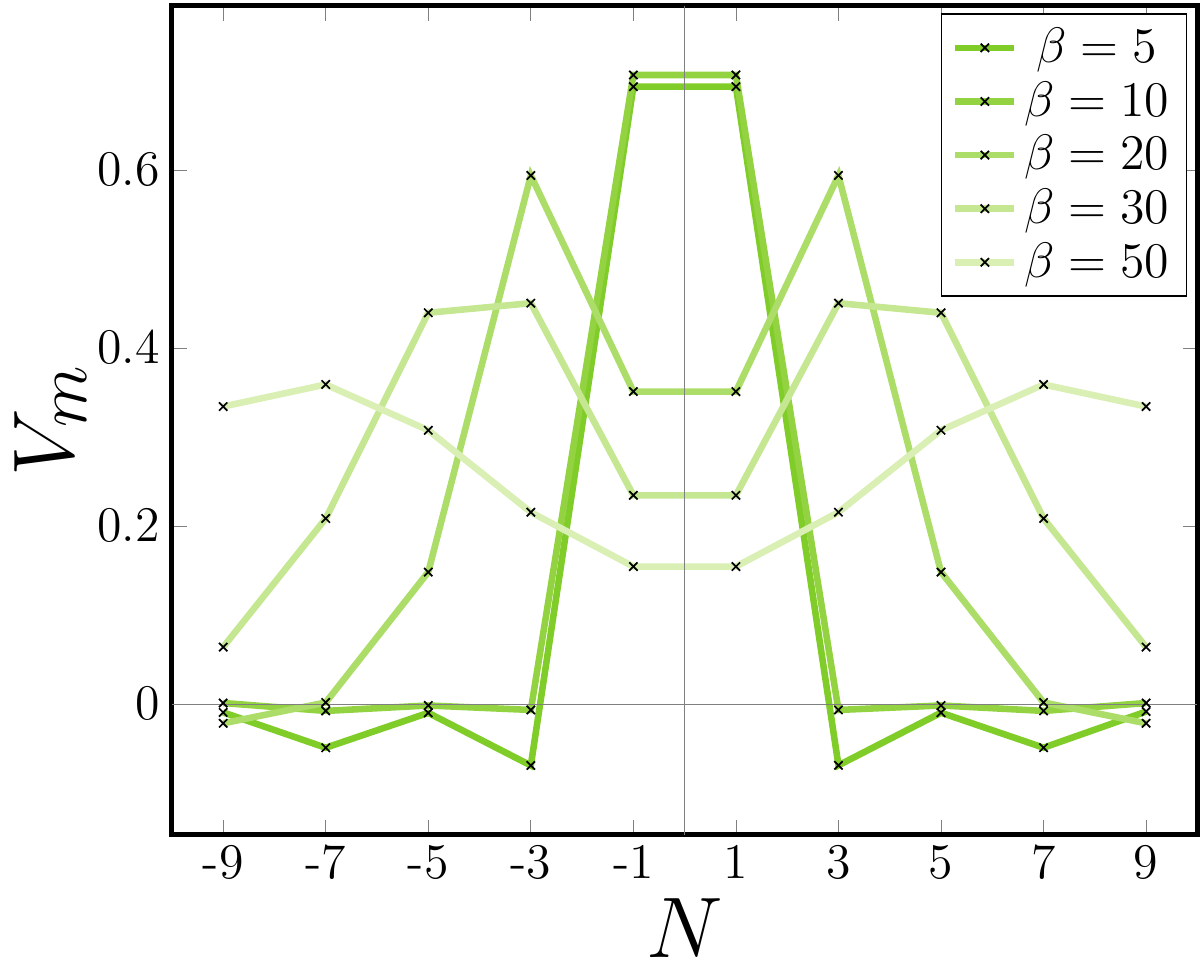}
\includegraphics[width=0.23\textwidth]{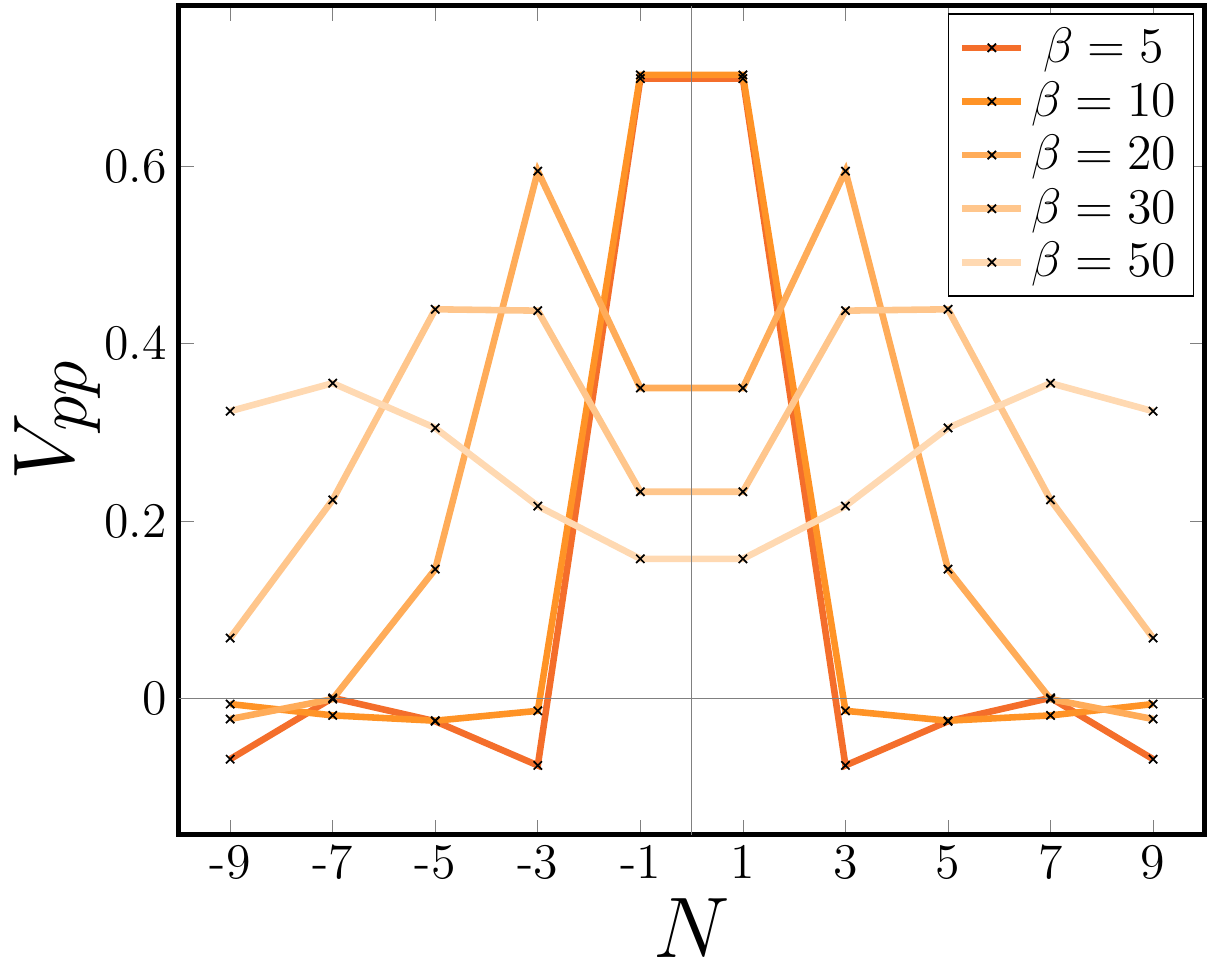}
\caption{ Comparison of the singular eigenvectors  $V^r_\alpha$ in the repulsive and the attractive case, plotted as a function of the Matsubara index $N=\nu \frac{\beta}{\pi}$.  The upper [lower] panel shows perfectly identical singular antisymmetric [symmetric] eigenvectors located at different temperatures along the first [second] attractive (left) and first [second] repulsive (right) divergence line.}
\label{Fig:2}
\end{figure}

When considering purely local quantities, the single-particle Green's function $G(\tau_1,\tau_2)$ is identical for the repulsive ($U\!>\!0$) and attractive ($U\!<\!0$) half-filled model. On the other hand, 
the two-particle Green's function $G_{\uparrow \downarrow}(\tau_1,\tau_2,\tau_3,\tau_4)$, i.e. the first time-ordered 
product appearing on the right hand side of Eq.\,\eqref{equ:form_gen_chi}, 
with anti-parallel spin orientation transforms\cite{Rohringer2012,Rohringer2013a} according to
\begin{equation}
G^{(U)}_{\uparrow \downarrow}(\tau_1,\tau_2,\tau_3,\tau_4) = -G^{(-U)}_{\uparrow \downarrow}(\tau_1,\tau_2,\tau_4,\tau_3),
\end{equation}
which, after Fourier transformation of all fermionic variables, reads
\begin{equation}
G^{(U)}_{\uparrow \downarrow}(\nu_1,\nu_2,\nu_3) = -G_{\uparrow \downarrow}^{(-U)}(\nu_1,\nu_2,-\nu_4) 
\end{equation}
with $\nu_4 = \nu_1 - \nu_2 + \nu_3$. After changing to the $ph$-notation, as defined in Eq.~\eqref{equ:form_gen_chi}, ($\nu_1=\nu$, $\nu_2=\nu+\Omega$, $\nu_3=\nu'+\Omega$, $\nu_4=\nu'$) one can easily see how the transformation maps the generalized static ($\Omega=0$) susceptibility, $\chi^{\nu,\nu'}_{\uparrow \downarrow}= G_{\uparrow \downarrow}(\nu, \nu, \nu')$ of the $\uparrow\downarrow$ sector according to
\begin{equation}
\label{eq:mapping_chi_U}
\chi^{\nu \nu'}_{\uparrow \downarrow}\,  \overset{U\leftrightarrow -U}{\Longleftrightarrow}  \, -\chi^{\nu (-\nu')}_{\uparrow \downarrow},
\end{equation} while  $\chi_{\uparrow \uparrow}$ is obviously invariant under a partial particle-hole transformation. 

Hence, in general, the Shiba transformation at the two-particle level  will mix the different (particle-hole) channels of 
generalized susceptibilities and the associated irreducible vertices.
 Only  the mapping of the generalized susceptibility expressed in the $pp$ notation 
\begin{equation}
\chi^{\nu (-\nu')}_{pp,\uparrow\downarrow} - \chi_{0,pp}^{\nu \nu'} \,  
\overset{U\leftrightarrow -U}{\Longleftrightarrow} \,   \chi^{\nu \nu'}_{m}
\label{eq:mapping_chi_pp}
\end{equation}
reflects\cite{Rohringer2012,Rohringer2013a,DelRe2019} the transformation of the physical (spin/pseudospin) degrees of 
freedoms, discussed in Eq.\,\eqref{Eq:SpinMap}, in a direct fashion.

As the location of divergence lines is
directly encoded in the generalized susceptibilities, it will be also subject to the mixing of channels, explaining the 
differences  w.r.t. the mapping of the physical degrees of freedom, discussed in Sec.\,\ref{sec:model}.
To fully rationalize the results observed in Sec.\,\ref{subsec:DMFTResults}, we will focus on the symmetry properties of the generalized susceptibilities. In this respect we note, that Eq.~(\ref{eq:mapping_chi_pp})
already shows why the divergences of the particle-particle $\uparrow\downarrow$ channel for $U>0$ are 
mirrored in the magnetic channel for $U<0$. Hence, the main question concerns the behavior of the 
particle-hole channels.

We start by considering the (spin resolved) generalized susceptibility $\chi^{\nu \nu' \Omega}_{\sigma \sigma'}$, as 
defined in Eq.~(\ref{equ:form_gen_chi}). Due to the particle-hole (PH) symmetry of the system considered here, $\chi^{\nu \nu' \Omega}_{\sigma \sigma'}$ has only real entries. Exploiting the time-reversal (TR)- and the SU(2)-symmetry 
of the problem \begin{eqnarray}
\big( \chi_{\sigma\sigma'}^{\nu \nu' \Omega} \big)^*  \overset{\scalebox{0.6}{PH}}{=} 
\chi_{\sigma\sigma'}^{\nu \nu' \Omega } 
\overset{\scalebox{0.6}{TR}}{=} \chi_{\sigma'\sigma}^{\nu'\nu \Omega}
\overset{\scalebox{0.6}{SU(2)}}{=} \chi_{\sigma\sigma'}^{\nu' \nu \Omega} \ .
\label{eq:chisym1}
\end{eqnarray}
it is evident that $\chi^{\nu \nu' \Omega}_{\sigma \sigma'}$ is a symmetric matrix of $\nu$ and $\nu'$. Relation \eqref{eq:chisym1} ensures that all matrix entries and all eigenvalues remain real for any $\Omega$.

Another symmetry relation can be obtained by exploiting the complex conjugation (CC) of $\chi^{\nu \nu' \Omega}_{\sigma \sigma'}$. For $\Omega\!=\!0$ it can be shown that the generalized susceptibility is invariant under the combined rotation of the matrix along both of its cardinal axes ($\nu\!\rightarrow\!-\nu, \nu'\!\rightarrow\!-\nu'$)
\begin{eqnarray}
\label{eq:symm_chi}
\chi_{\sigma\sigma'}^{\nu \nu'} \overset{\scalebox{0.6}{PH}}{=} \big( \chi_{\sigma\sigma'}^{\nu \nu'} \big)^*  
&\overset{\scalebox{0.6}{CC}}{=}& {\chi_{\sigma'\sigma}^{(-\nu') (-\nu)} }
\overset{\scalebox{0.6}{TR}}{=} {\chi_{\sigma\sigma'}^{(-\nu) (-\nu')}} \ .
\end{eqnarray}
A matrix obeying the conditions 
\begin{eqnarray}
\label{eq:centrosymm_chi}
\chi_{\sigma\sigma'}^{\nu \nu'} = \chi_{\sigma\sigma'}^{(-\nu) (-\nu')} \quad \text{and} \quad \chi_{\sigma\sigma'}^{\nu \nu' } = \chi_{\sigma\sigma'}^{\nu' \nu} 
\end{eqnarray}
is a so-called {\sl bisymmetric} matrix, where the matrix elements are symmetric with respect to both the main diagonal ($\nu = \nu'$) as well as the secondary diagonal ($\nu=-\nu'$). Essentially, this particular symmetry is at the core to understand the mapping of divergence lines. 

A bisymmetric matrix can always be diagonalized into blocks (here associated to positive/negative Matsubara frequencies), by applying an orthogonal matrix $Q$,  defined  in terms of the counteridentity $(J^{\nu,\nu'}\!=\!\delta_{\nu (-\nu')}$) and identity submatrices $\mathbb{1}$ (see Appendix~\ref{Appendix:BlockDiag} for more details)
\begin{equation}
Q=\frac{1}{\sqrt{2}}
\begin{pmatrix}
\mathbb{1}  & -J \\
\mathbb{1} & J 
\end{pmatrix}
\quad , \quad 
Q\chi_rQ^T =
\begin{pmatrix}
\text{A} & \vline & 0 \\
\hline
0 & \vline & \text{S} \\
\end{pmatrix} \ .
\label{Eq:Subspaces}
\end{equation}
The block-diagonalization of $\chi_r$ is associated with precise symmetry properties: the subspace denoted by A represents a submatrix with exclusively antisymmetric eigenvectors, while S is the subspace of purely symmetric eigenvectors. 
As a consequence, one can unambiguously attribute the occurrence of a red divergence line in $\chi_d$ to the purely antisymmetric subspace A, while all other divergence lines will be accounted for by the symmetric subspace S.

\begin{figure*}[hbt!]
\includegraphics[width=0.43\textwidth,height=0.33\textwidth]{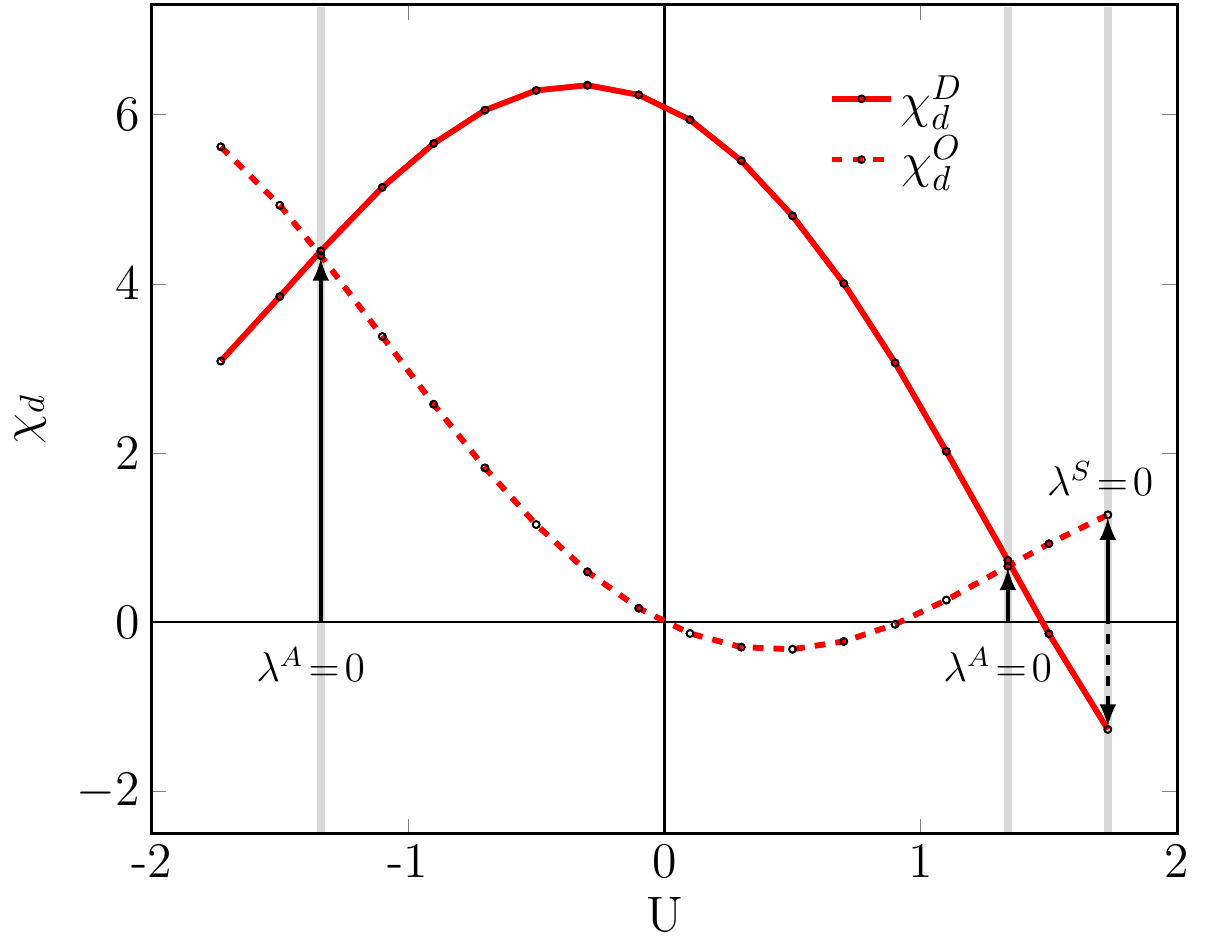} \hspace{5mm}
\includegraphics[width=0.43\textwidth,height=0.33\textwidth ]{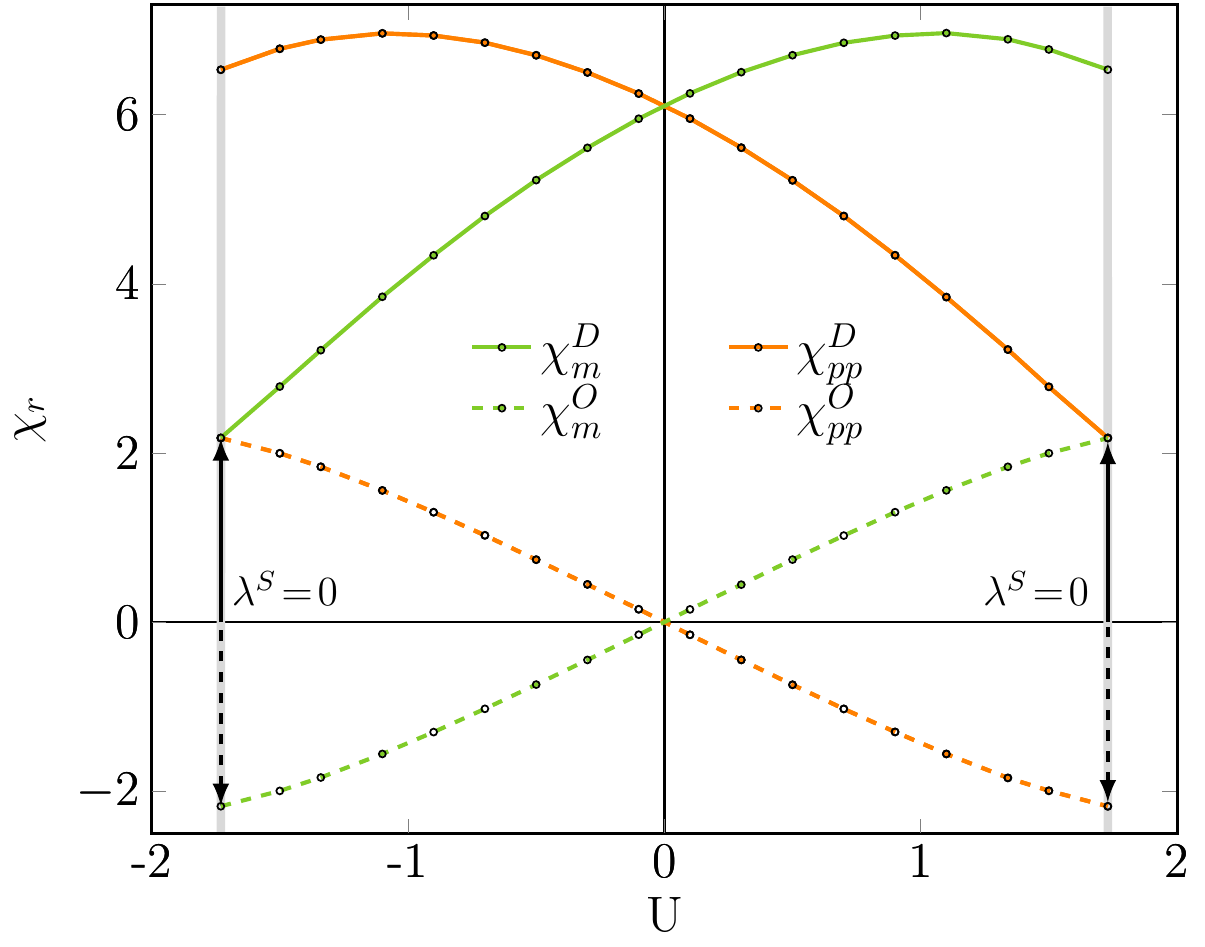}
\caption{Evolution of the diagonal ($D$) and the off-diagonal ($O$) innermost matrix elements of the generalized susceptibilities, computed in DMFT, as a function of the local (attractive and repulsive) Hubbard interaction (left panel: density channel; right panel: magnetic and pp channels)  in the high-T regime ($\beta =5$). 
The  vertical  bars in light grey mark the U values of  the vertex divergences in the corresponding sectors, while the black arrows highlight the fulfillment of the singularity conditions for the $2 \! \times 2$ innermost frequency matrix of  $\chi^{\nu,\nu'}_r(\Omega)$, namely  $\chi_r^D\! = \!\chi_r^O$  (associated to antisymmetric singular eigenvectors) and  $\chi_r^D\! =  -\!\chi_r^O$  (symmetric eigenvectors).}
\label{Fig:MatElTrendb5}
\end{figure*}


A crucial ingredient for connecting the bisymmetry of the generalized susceptibilities to the mapping of divergence lines 
lies in the equivalence of the Shiba transformation for $\chi^{\nu\nu'}_{\uparrow \downarrow}$ 
to a matrix multiplication with the negative counteridentity matrix $(-J)$ 
\begin{eqnarray}
\label{eq:mapping_chi_U}
	\chi^{\nu\nu'}_{\uparrow \downarrow,(U)}(-J) = 
 -\chi^{\nu(-\nu')}_{\uparrow \downarrow,(U)} = \chi^{\nu\nu'}_{\uparrow \downarrow,(-U)} \ .
 \label{eq:JMap}
\end{eqnarray}
Combining Eq.~\eqref{Eq:Subspaces}, \eqref{eq:JMap} and the fact that $J^2=\mathbb{1}$ one can prove (see 
Appendix~\ref{App:equality}) the remarkable result, that the antisymmetric sector A remains {\sl invariant} under $U\!\leftrightarrow\!-U$ for all $\chi_r$.
This explains 
why the red divergence lines ($\chi_d$) in Fig.\,\ref{Fig:1} and their associated antisymmetric eigenvectors (Fig.~\ref{Fig:2})
are perfectly mirrored on both sides of the phase-diagram.
At the same time one finds that the symmetric parts (S) of $\chi_d$ and $\chi_m$ are mapped into one-another for $U\!\leftrightarrow\!-U$, therefore connecting the symmetric divergences and the corresponding eigenvectors, appearing in $\chi_d^{U>0}$ (orange) and in $\chi_m^{U<0}$ (green).

Let us stress that the proof given in 
Appendix~\ref{App:equality} applies not only to singular eigenvalues, which are
connected to divergence lines, but to all eigenvalues and eigenvectors of $\chi^{\nu\nu'}_{r}$. 
In this way, we have extended the mapping relation known for $\chi^{\nu\nu'}_{pp,\uparrow\downarrow}$
to the entire particle-hole sector, clarifying the relation with the mapping of the physical D.o.F.: 
the  antisymmetric subspace A, not contributing to the sum for the physical susceptibility in Eq.\,\eqref{eq:ChiSpectrum}, is {\sl  invariant} under the Shiba transformation, while the symmetric subspace is found to transform in accordance with Eq.\,\eqref{Eq:SpinMap}. 

As we have illustrated, the particle-hole symmetry plays a central role in determining the mirroring 
properties of the generalized susceptibilities. If one relaxes this constraint, 
the relations in Eq.~\eqref{eq:centrosymm_chi} no longer hold in the particle-hole sector. Therefore, the bisymmetry is lost and eigenvalues are not necessarily real. 
This implies that the eigenvectors of the corresponding $\chi_r$ are not necessarily symmetric or antisymmetric any longer. 
At the same time, it is important to stress, that even in the absence of PH-symmetry (e.g. out of half-filling) 
$\chi^{\nu \nu'}_{pp, \uparrow\downarrow}$ continues to fulfill\cite{Thunstroem2018} both relations in Eq. \eqref{eq:centrosymm_chi}, ensuring the validity of all associated properties (real eigenvalues as well as bisymmetry and associated properties).

\subsection{High-Temperature Limit}

To exemplify the concepts discussed in the previous section, we performed  DMFT calculations in the high-temperature regime ($\beta\!=\!5$), where the frequency structure of the two-particle generalized susceptibilities strongly simplifies. Due to the large step size on the Matsubara frequency grid, most information of the system is encoded in the central $2\times2$ matrix. 
The analysis of the divergences can then be restricted\cite{Gunnarsson2016} to the innermost $2\!  \times  \! 2 $ matrix defined by the smallest Matsubara frequencies ($\nu, \nu'= - \frac{\pi}{\beta}, \frac{\pi}{\beta}$).

For a $2 \times 2$ case, the bisymmetry condition (see III B) poses significant constraints on the matrix elements and a singularity can be realized only in two ways:
\begin{eqnarray}
  \chi_r^{\lambda_A=0}&=&
  \left(\begin{array}{cccc}
     \colorbox{red!15}{\color{black!100}{a}} & \colorbox{red!15}{\color{black!100}{a}} \\
     \colorbox{red!15}{\color{black!100}{a}} & \colorbox{red!15}{\color{black!100}{a}} \\
  \end{array}\right)
\label{Eq:Centro1}
\end{eqnarray}
which corresponds to the (anti-symmetric) singular eigenvector $V_A (\nu) \propto \, \delta_{\nu,\frac{\pi}{\beta}} - \delta_{\nu, -\frac{\pi}{\beta}}$, and
\begin{eqnarray}
  \chi_r^{\lambda_S=0}&=&
  \left(\begin{array}{cccc}
     \colorbox{blue!15}{\color{black}{$\mp b$}} & \colorbox{red!15}{\color{black}{$\pm b$}}  \\
     \colorbox{red!15}{\color{black}{$\pm b$}} & \colorbox{blue!15}{\color{black}{$\mp b$}}  \\
  \end{array}\right) 
  \scriptsize
  \normalsize
\label{Eq:Centro2}
\end{eqnarray}
with  $a, b> 0$, corresponding to a (symmetric) singular eigenvector   $V_S  (\nu) \propto \, \delta_{\nu,\frac{\pi}{\beta}} + \delta_{\nu, -\frac{\pi}{\beta}}$.  

On the basis of these considerations, we analyze the $U$-dependence of the diagonal ($\chi_r^D$) and off-diagonal ($\chi_r^O$) elements of the $2\!\times\!2$ lowest frequency-submatrix of the generalized susceptibility, extending the study of Ref.~[\onlinecite{Gunnarsson2016}] to the attractive case. 
The corresponding data are reported in Fig.~\ref{Fig:MatElTrendb5} for the density (left) and the magnetic/pp sectors (right).

A general trend can readily be identified: Upon increasing $|U|$ all diagonal matrix elements ($\chi_r^D$) eventually decrease, while the off-diagonal elements ($\chi_r^O$) mostly increase in absolute values, for the considered interaction regime. 
The decrease of $\chi_r^D$ upon increasing $|U|$ is dominated by the bubble term ($\propto - \beta \,G(\nu)G(\nu') \, \delta_{\nu \nu'}$), reflecting the suppression of the single particle Green's function $G(\nu)$ at low-frequencies. 
Vertex corrections are responsible for the asymmetry of the damping effects on $\chi_r^D$ with respect to $\pm U$ as well as for its different size in the different sectors.
 
In particular, we find the following behavior for the diagonal entries: (i) the decrease-rate with $|U|$ of $\chi_r^D$ is stronger in those channels that correspond to a suppressed susceptibility, (ii)  $\chi_d^D$ decreases faster compared to the other two channels and even turns negative for large $U>0$, where density fluctuations are suppressed.
 
The off-diagonal matrix elements are obviously zero in the non-interacting case ($U=0$) and for small values of $U$ yield positive/negative corrections to the enhanced/suppressed susceptibilities. For large $U$ values this behavior is preserved in the $m$ and $pp$ channel. An exception is the suppressed density channel where $\chi_d^O$ displays a strong increase, becoming positive again.
 
From these observations, we conclude that the suppression/enhancement of a static physical susceptibility is controlled by the interplay of suppressed diagonal entries and the enhanced magnitude of the (positive/negative) off-diagonal terms. 
 
Due to the considerably milder damping of the diagonal entries in the magnetic and the $pp$ sector, one always finds that $\chi_r^D > \chi_r^O$, for $r=m,pp$. Therefore, only singularities of the second kind ($\chi_r^D =  - \chi_r^O$, s.\ Eq.~(\ref{Eq:Centro2})) can occur in these channels. This implies that singularities of the second kind can occur exclusively in sectors of suppressed susceptibilities. 
 
On the contrary, the much stronger damping of $\chi_d^D$ plays a crucial role in suppressing the density fluctuations for $U>0$. For $U<0$ this decrease of $\chi_d^D$ is outperformed by an even stronger increase of $\chi_d^O$ in order to describe the corresponding enhancement of $\chi_d$. As one can easily  see in Fig.~\ref{Fig:MatElTrendb5}, these conditions allow divergences of the first kind with $\chi_d^D = \chi_d^O$ 
(compare Eq.~\eqref{Eq:Centro1}), to occur specularly on both sides of the  phase-diagram. In fact, frequency-antisymmetric divergences are the only one to be expected in sectors of enhanced physical susceptibilities, because in this regime, both  diagonal and off-diagonal components of  $\chi_r^{\nu,\nu'}$  have the same (positive) sign.

\section{Spectral representations of physical susceptibilities}
\label{sec:spectral}
\noindent

The relation between generalized susceptibilities (and associated vertex divergences) and the physical response of our system can be illustrated in a more insightful way, by exploiting the following spectral representation. 

As all eigenvalues in Eq. (\ref{eq:ChiSpectrum}) are real, we introduce a susceptibility density ($\rho(\chi)$) defined as
\begin{equation}
\rho_r(\chi) = \sum_\ell \left\vert \sum_{\nu} V^r_\ell(\nu) \right\vert^2 \delta(\chi - \lambda^r_\ell) \geq 0
\label{e1}
\end{equation}
from which the local physical susceptibility is readily obtained as an average over $\rho_r(\chi)$
\begin{equation}
\langle \chi_r \rangle = \int \chi \, \,  \rho_r(\chi) \, d\chi \ . 
\label{eq2}
\end{equation}
This representation, inspired from the analysis of Ref.~\onlinecite{Gunnarsson2017}, has several advantages:  Equations (\ref{e1}) and (\ref{eq2}) enable to distinguish between positive ($\lambda^\ell>0, \rho(\lambda^\ell)>0$), negative ($\lambda^\ell<0, \rho(\lambda^\ell)>0$) and vanishing ($\lambda^\ell=0$ or $\rho(\lambda^\ell)=0$) contributions to the static response $\chi_r$. Further, its graphical conciseness is particularly suited to illustrate how the mapping of the generalized susceptibilities works for the different cases, highlighting the most relevant physical implications.

\begin{figure*}[t]
\includegraphics[width=0.32\textwidth]{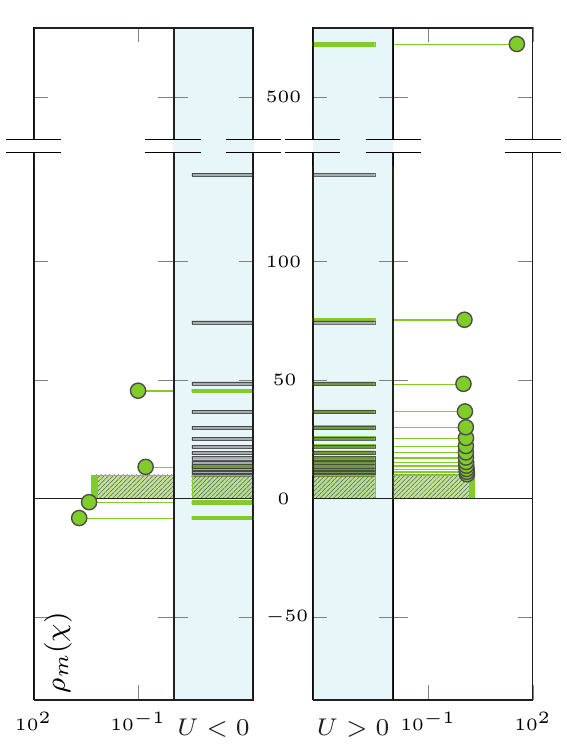}
\includegraphics[width=0.32\textwidth]{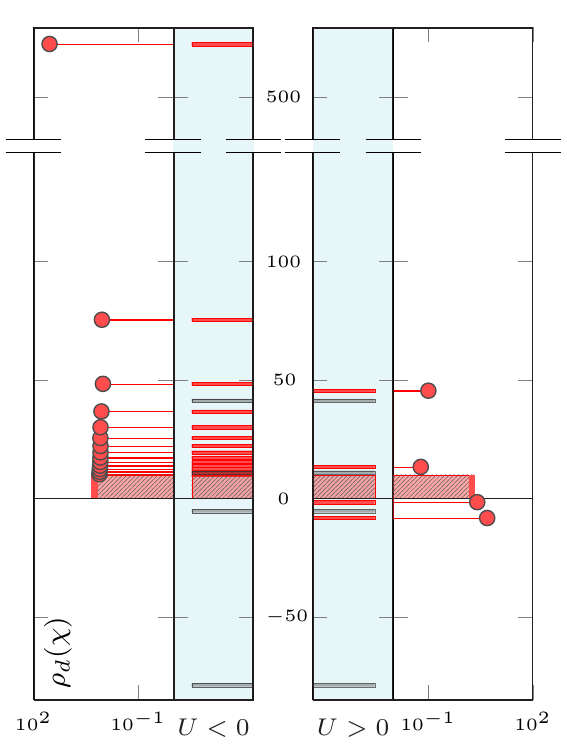}
\includegraphics[width=0.32\textwidth]{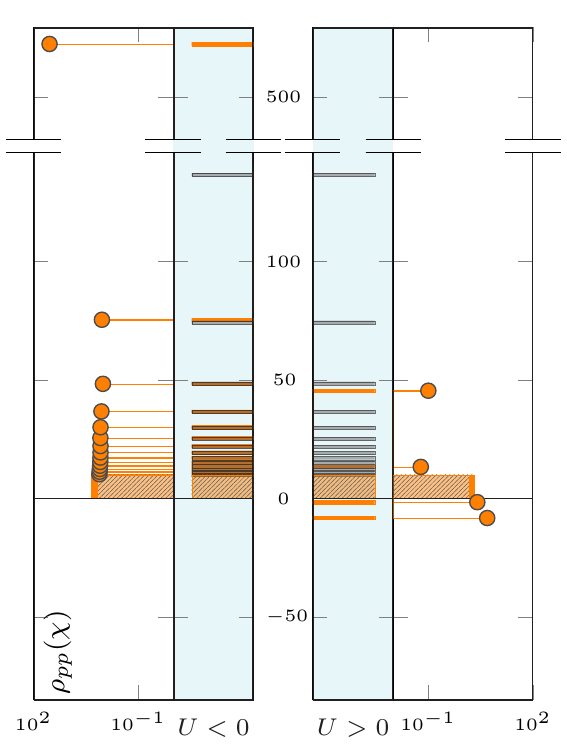}
\caption{Comparison of susceptibility densities Eq.~\eqref{eq2}  for the magnetic (green), density (red), and $pp$ (orange) sectors of the attractive $U<0$ and repulsive $U>0$ case, respectively. Each $\delta$ function in Eq.~(\ref{eq2}) is represented by horizontal bars (grey/colored if corresponding to an antisymmetric/symmetric EV). The corresponding densities $\rho_r(\chi)$ are plotted as colored circles. Data shown here was obtained for $T=0.2$ and $U=2.16$. The position in the phase diagram (Fig.~\ref{Fig:1}) is indicated by light blue stars .}
\label{Fig:Densities}
\end{figure*}

The introduced representation is applied here to analyze our susceptibility data after crossing four divergence lines at two mirrored positions in the phase diagram (light-blue stars in Fig.\ref{Fig:1}).
 
The corresponding results are shown in the three plots of Fig.~\ref{Fig:Densities}, representing the three scattering channels.
The positions of all eigenvalues $\lambda_\ell^r$ are shown as bars in the light-blue shaded innermost panels of the three plots: 
Gray bars indicate eigenvalues associated to antisymmetric eigenvectors and thus to a vanishing $\rho_r$ which does not contribute to $\chi_r$.  Colored bars account for eigenvalues associated to finite $\rho_r(\lambda_\ell^r)$ values, corresponding to symmetric eigenvectors whose weighted sum builds up the full $\chi_r$. 
The actual value of   the susceptibility-density $\rho_r$ for a given eigenvalue is indicated by the circle-symbols in the outermost panels of the plots in Fig.~\ref{Fig:Densities}. The color-shaded regions slightly above $\chi\!\sim \! 0$ represent an increasingly denser distribution of small positive eigenvalues, arising from the high-frequency behavior of  $\chi_r^{\nu \nu} \propto \, \frac{1}{\nu^2} \, \delta^{\nu\nu'}$. It can be shown that this (essentially non-interacting) large-$\nu$ feature induces a van Hove singularity in the $T \rightarrow 0$ behavior of $\rho_r(\chi) \simeq 1/\chi^{-3/2}$  for $\chi \rightarrow 0$ (See Appendix~\ref{App:BM}).

The three plots of  Fig.~\ref{Fig:Densities} graphically combine  all aspects of the attractive-repulsive mapping of the generalized susceptibilities and allow a comprehensive understanding at a single glance.

The location of the colored bars together with the corresponding values of $\rho_r(\chi)$ are transformed fully consistent with the mapping of the physical D.o.F.. In accordance with our results in Sec. IIIB, not only the physical susceptibility, but the entire distribution $\rho_r(\chi)$ of the identical density and $pp$ (pseudospin) sectors are mapped onto the magnetic (spin) sector and vice versa. 

On the contrary, the positions of the gray bars of each channel are unchanged in the $+U$ and $-U$ cases, reflecting the  {\sl invariance} of the antisymmetric subspaces of all generalized $\chi_r$ under the mapping. We note that the identical location of the gray bars in the magnetic and the $pp$ channel 
reflects the fact that the entire generalized susceptibility sectors are transformed exactly as the physical degrees of freedom (compare Eq.~\eqref{eq:mapping_chi_pp}).

On the other hand, the different locations of gray bars in the 
density sector compared to the other channels explain the non-trivial mapping properties of $\chi_{d}^{\nu\nu'}$ and of the corresponding irreducible vertices.

These general observations allow for a remarkable rationalization: Any suppressed local physical susceptibility can be associated to a  {\sl unique} susceptibility-density 
\begin{equation}
\rho_{sup} (U)\!=\!\rho_{m}^{U<0}\!=\!\rho_{d}^{U>0}\!=\!\rho_{pp}^{U>0} \ .
\label{eq:rhosup}
\end{equation}
Obviously, by replacing $U$ with $-U$ in Eq.~\eqref{eq:rhosup}, a similar property holds for all enhanced susceptibility densities
\begin{equation}
\rho_{enh}\!(U)= \rho_{sup}(-U) =\!\rho_{m}^{U>0}\!=\!\rho_{d}^{U<0}\!=\!\rho_{pp}^{U<0} \ .
\label{eq:rhoenh}
\end{equation}

The comparison of the attractive and repulsive panels of each channel in Fig.~\ref{Fig:Densities} indicates as an overall trend, that the suppression of a susceptibility is associated to a systematic shift of the colored bars towards smaller values, as well as with a change of the weight distribution, where the largest values of $\rho_{sup}$ are associated with the smallest eigenvalues.
This supports the physical picture that an interaction-driven suppression of a static local susceptibility is connected to an increasing number of negative eigenvalues and therefore with the crossing of multiple vertex divergences. This corresponds to a loose generalization of the self-energy behavior at the 2P level, as discussed in Sec.~\ref{sec:model}. 

At the same time, this demonstrates why the reverse implication of the above physical picture is not correct. The invariance of the gray bars under the mapping implies the mirroring of all red vertex divergence lines, where only the density channel is singular (Fig.~\ref{Fig:1}). Hence, the occurrence of red divergence lines is  independent of the behavior of the corresponding susceptibility as well as of the SU(2)$\times$SU(2) symmetry properties of the model considered. 

\begin{figure}[t]
\includegraphics[width=0.45\textwidth]{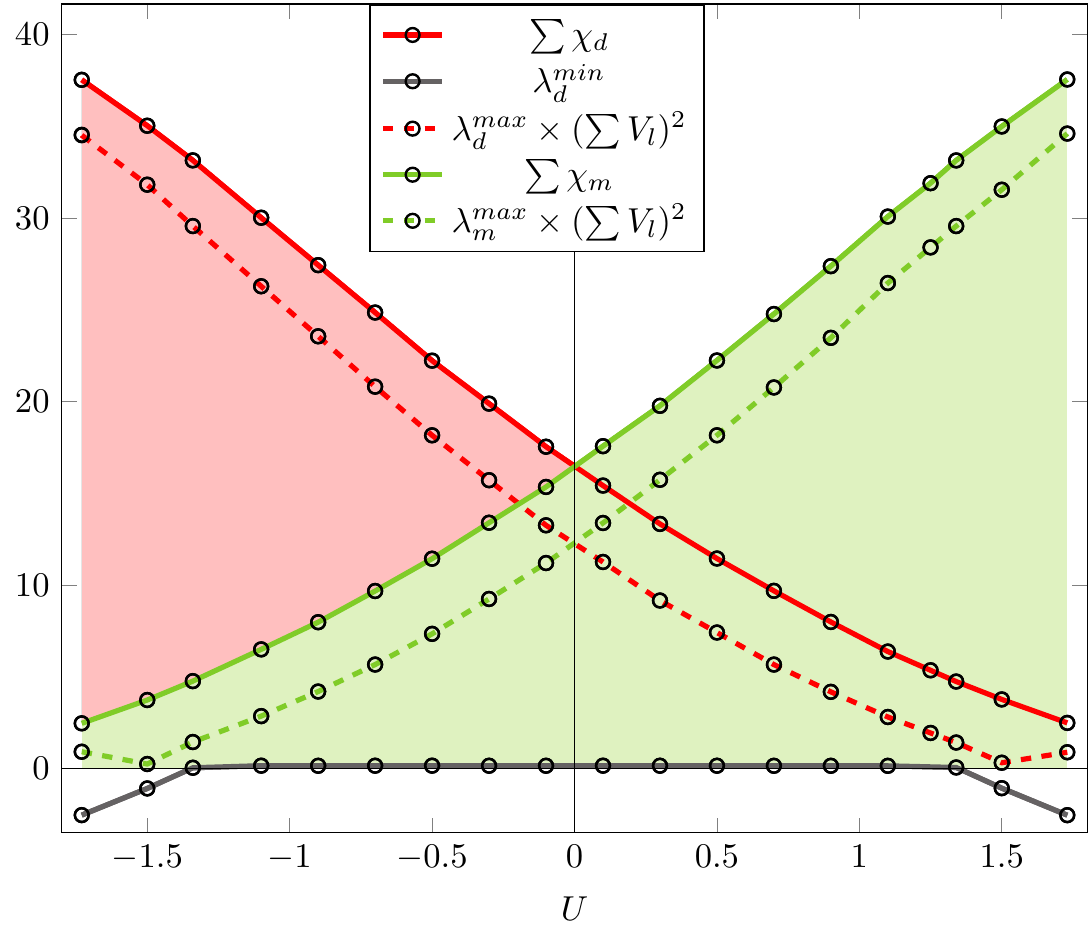}
\caption{Comparison of the static density $\chi_d(\Omega=0)$ (red) and magnetic $\chi_m(\Omega=0)$ (green) susceptibility with the contribution of the largest eigenvalue only, as a function of the attractive/repulsive Hubbard interaction $U$ at $T=0.2$. In the bottom of the plot, the lowest eigenvalue of $\chi_d^{\nu \nu'}$ is shown. The evolution of the lowest eigenvalues ($\lambda^{min}_d$, in dark gray) is completely decoupled from the behavior of the
static susceptibility.}
\label{Fig:ChargeTrend}
\end{figure}

Finally, important quantitative information can be also gained from Fig.~\ref{Fig:Densities}.
By analyzing the behavior of the enhanced susceptibilities, it is evident that $\rho_{enh}$ is dominated by the contribution of 
a  single term: the one associated to largest eigenvalue $\lambda^{max}$. This property is illustrated in Fig.~\ref{Fig:ChargeTrend}, where we compare the actual values of $\chi_d $ and $\chi_m$ obtained from Eq.~\eqref{eq2} with the case where the summation in Eq.~\eqref{e1} is reduced to the largest eigenvalue only. The contribution from the largest eigenvalue $\lambda_{max}$ very well reproduces the trend across the entire repulsive and attractive regime and even well approximates the actual value of the static susceptibilities $ \chi_d$ and $\chi_m$ in their respective enhanced regions. 
Since the relation $V_{enh}^{max}=V_m^{max}=V_{d}^{max}=V_{pp}^{max}$ follows from the proof in Appendix~\ref{App:equality} and
Eq.~\ref{eq:mapping_chi_pp}, the value of all physical susceptibilities in their respective enhanced regions can be well approximated by
\begin{equation}
\langle \chi_{r} \rangle \sim \lambda^{max} \left\vert \sum_{\nu} V^{max}_{enh}(\nu)  \right\vert^2\ . 
\end{equation}
According to this relation, the Curie-Weiss behavior of any static local susceptibility in the strong-coupling regime can be ascribed to the evolution of the corresponding $\lambda^{max}$ and the associated eigenvector. 

\section{Physical and algorithmic consequences}
\label{sec:consequences}

The results presented in the previous sections allow us to make some considerations on the possible implications of vertex divergences.

 Our DMFT study has proven that vertex divergences associated with antisymmetric singular eigenvectors  can  affect also the dominant scattering channel. This observation is relevant for the usage of parquet-based schemes in the non perturbative regime, such as D$\Gamma$A\cite{Toschi2007} and QUADRILEX\cite{Ayral2016}.
In fact, in all cases where vertex divergences are confined to the secondary scattering channels, their appearance can be exploited as an useful indicator that these sectors can be safely neglected. This would considerably simplify the parquet treatment of the problem under investigation (e.g. reducing the parquet treatment to an effective BSE-based algorithm and avoiding the explicit manipulation of divergent quantities). Unfortunately, the occurrence of divergences in the dominant channels prevents a straightforward implementation of this idea. Hence, other ways to handle parquet equations at strong-coupling must be followed, such as the combination of fRG and DMFT, 
($\text{DMF}^2\text{RG}$) \cite{Taranto2014} or the single-boson exchange (SBE) approach\cite{Krien2019SBE}.

At the same time, we note that the antisymmetric nature of the divergences occurring in the dominant channels will not hinder the applicability of post-processing schemes of non-perturbative results based on the parquet equations (e.g.\ the parquet-decomposition of the self-energy\cite{Gunnarsson2016}), since the potentially dangerous effects of such divergences will be cancelled by the internal summation over fermionic variables. 
Even if this is more speculative, we think that the observation above might inspire alternative strategies to circumvent the divergences occurring in the major channels, even at the level of parquet solvers\cite{Tam2013,Valli2015,Wentzell2016,Li2016,Kauch2019}, by exploiting the odd symmetry properties of their frequencies (and/or momentum\cite{Gunnarsson2016}) structures. 
We should also note that the divergences associated to antisymmetric eigenvectors are the first to be encountered upon increasing the interaction, independent of the interaction sign. As they affect the density channel, it is plausible to expect that diagrammatic Monte Carlo algorithms based on  bold resummations are going to encounter similar difficulties of formal convergence towards unphysical solutions both for repulsive\cite{Kozik2015} and attractive interactions. 

Further, we recall that our results (and in particular that of Sec. III B and Sec. IV) do not only apply to the singular eigenvalues and eigenvectors.
Instead, they fully define the effects of the Shiba mapping on all generalized two-particle quantities: The symmetric subspaces of $\chi_r$ are transformed exactly in the same way as the physical D.o.F., while the antisymmetric subspaces remain {\sl invariant}.

Hence, consistently with the total decoupling of the antisymmetric eigenvectors from the static susceptibilities (see Secs. IIIB and IV), one would be tempted to associate the whole physically relevant information with the symmetric subspace of the generalized susceptibilities.  However, this is not valid in general. In fact, while the antisymmetric subspace of $\chi_{r}$ does not contribute at all to the corresponding static susceptibility, it can affect the behavior of other physical quantities.  Pertinent examples are energy-energy correlation-functions, i.e. response functions which explicitly contain first-order time-derivatives (i.e., $i\hbar \frac{d}{d t} = - \hbar \frac{d}{d \tau} = \hat{H}$ and, hence, odd function of $\nu$, $'\nu'$, after Fourier transform) such as the thermal conductivity\footnote{Note that the thermal conductivity expression contains also the spatial derivatives of the heat-current. This prevents any kind of vertex-corrections in the single-orbital model at the DMFT level\cite{Georges1996}. The latter are possible, however, at the level of DCA\cite{Maier2005}, where vertex-divergences with antisymmetric structure have been also reported\cite{Gunnarsson2016}. At the same time, correlation functions containing time-derivatives, where the effects of the antisymmetric part of the vertex functions are important, could be the most suited for establishing a connection with non-equilibrium phenomena.}.


Finally, we note that if symmetries of the problem are lifted (e.g. by doping the system, considering further hopping terms or applying a magnetic field, etc.), corresponding changes must be expected\cite{Vucicevic2018,Reitner2020,vanLoon2020}. In this respect, the high-symmetry case we considered in this work will represent a good compass for interpreting the deviations which will be observed in future studies.

\section{Conclusion}

In the present work we conducted a comparative DMFT analysis to understand the location and physical role of vertex divergences occurring in the two-particle vertex correlation functions of the repulsive and attractive Hubbard model. Our DMFT calculations show that the location of divergences of two-particle irreducible vertices is perfectly symmetric in the attractive and repulsive Hubbard model. 

This result partly contradicts the expectation from the one-particle picture, where a divergence of the self-energy is accompanied by the suppression of the one-particle Green's function. In particular the symmetric occurrence of singular eigenvalues in $\chi^{\nu \nu'}_d$ for $U\!\lessgtr \! 0$ shows, that divergences of the two-particle self-energy $\Gamma$ do not necessarily occur in physically suppressed channels. 

A thorough interpretation of our numerical results has been gained by analyzing the specific symmetries that apply in the presently considered system.  In particular,  we show that the antisymmetric and symmetric subspaces behave differently under the $U\!\leftrightarrow\!-U$ transformation. The antisymmetric part of the generalized susceptibilities is {\sl invariant} under the Shiba transformation, hence explaining the perfectly mirrored red divergence lines in the density sector, while for the symmetric subspace on the other hand, the density, particle-particle and magnetic channels are mapped into each other for $U\leftrightarrow-U$. 

Therefore, we confirm that the interaction-driven suppression of a static local susceptibility is generally accompanied by an increasing number of negative eigenvalues,  if they are associated to symmetric eigenvectors, which actively contribute to the suppression of the channel. However, the reversed implication, that the occurrence of negative eigenvalues is in general indicative of the suppression of a channel, is  not valid, because of the antisymmetric divergence lines being invariant under $U\leftrightarrow -U$. 

This suggests to represent the physically relevant information in terms of a susceptibility density distribution which naturally distinguish the symmetric from the vanishing antisymmetric eigenvector subspace. This representation allows to summarize the $U\!\leftrightarrow\!-U$ mapping behavior of the generalized susceptibilities and its relation to the mapping of the physical (spin and pseudospin) degrees of freedom at a single glance. Moreover, since the associated spectral distribution is identical for all suppressed as well as all enhanced channels, the introduced representation provides a universal description of  all physical susceptibilities relevant for this problem.

Further studies are required to clarify the role of the  antisymmetric subspace for other physical quantities such as the thermal conductivity, the effect of a progressive reduction of the symmetry conditions and, on a broader perspective, the relation with the non-equilibrium properties of the system under investigation.


\vskip 5mm

{\sl Acknowledgments:} We are indebted for insightful discussions with  Sabine Andergassen, Massimo Capone, Lorenzo Del Re, James Freericks, Anna Kauch, Olle Gunnarsson,  Andreas Hausoel, Cornelia Hille, Friedrich Krien, Erik van Loon, Matthias Reitner, Georg Rohringer, Thomas Sch\"afer, Agnese Tagliavini, Patrik Thunstr\"om, and  Angelo Valli. We acknowledge financial support from the Austrian Science Fund (FWF) through the projects:  SFB ViCoM F41 (DS) and I 2794-N35  (PC, AT). Calculations have been performed on Vienna Scientific Cluster (VSC).

\appendix
\section{Bisymmetric Matrices}
\label{App:CentrosymmetricMatrices}

The following part is a short summary of mathematical literature on 
bisymmetric and centrosymmetric matrices\cite{CentrosymmMatrix1, CentrosymmMatrix2, CentrosymmMatrix3},necessary to follow the proof in part~\ref{App:equality}.

Note that at this point we focus on the matrix properties related to Eq.~\eqref{eq:centrosymm_chi}, 
without taking into account that the matrix is also symmetric. In this case one speaks of 
{\sl centrosymmetric} matrices. 

In the following we consider a centrosymmetric matrix $H$, a $2n\times 2n$ matrix, where $n$ is the number of positive/negative fermionic Matsubara frequencies. 
As $H$ is a centrosymmetric matrix it fulfills the following condition

\begin{equation}
\label{equ:def}
JHJ=H
\end{equation}
where $J$ is the counteridentity matrix ($J^2=\mathbb{1}$) defined as 

\begin{equation}
\label{equ:appix_centrosymm_J}
J=
\begin{pmatrix}
0 & \dots & 0 & 1 \\
\vdots &\reflectbox{$\ddots$} & \reflectbox{$\ddots$} & 0 \\
0 & 1 & \reflectbox{$\ddots$} & \vdots \\
1 & 0 & \dots & 0
\end{pmatrix}
=
\begin{pmatrix}
\mathbb{0} & J \\
J & \mathbb{0}
\end{pmatrix} \ .
\end{equation}
If $J$ is multiplied from the right, it 
inverts the columns of a matrix, if it is multiplied from the left, the rows are inverted. 
As one can easily see, for $\chi_{\sigma\sigma'}^{\nu \nu'}$ this implies

\begin{equation}
J\chi_{\sigma\sigma'}^{\nu \nu'}J=J\chi_{\sigma\sigma'}^{\nu (-\nu')} = \chi_{\sigma\sigma'}^{(-\nu) (-\nu')} = \chi_{\sigma\sigma'}^{\nu \nu'} \, ,
\end{equation}
which is true for our case, see Eq.~(\ref{eq:centrosymm_chi}) in the main text.

If H is a centrosymmetric matrix, the following condition holds, where 
the submatrices $A,B,C,D$ are $n\times n$ matrices

\begin{eqnarray}
H=
\begin{pmatrix}
A & \vline & B \\
\hline 
C & \vline & D 
\end{pmatrix}
&\stackrel{(\ref{equ:def})}{=}&JHJ \nonumber
\\
\begin{pmatrix} 
A & \vline & B \\
\hline 
C & \vline & D 
\end{pmatrix} 
&=&
\begin{pmatrix}
\mathbb{0} & J \\
J & \mathbb{0} 
\end{pmatrix}
\begin{pmatrix}
A & \vline & B \\
\hline 
C & \vline & D 
\end{pmatrix}
\begin{pmatrix}
\mathbb{0} & J \\
J & \mathbb{0} 
\end{pmatrix} \nonumber
\\
&=&
\begin{pmatrix}
\mathbb{0} & J \\
J & \mathbb{0} 
\end{pmatrix}
\begin{pmatrix}
BJ & \vline & AJ \\
\hline 
DJ & \vline & CJ 
\end{pmatrix} \nonumber
\\
&=&
\begin{pmatrix}
JDJ & \vline & JCJ \\
\hline 
JBJ & \vline & JAJ 
\end{pmatrix}
\end{eqnarray}

\begin{equation}
\Rightarrow D=JAJ \quad\&\quad B=JCJ \ .
\end{equation}
This means that the centrosymmetric matrix H can be written in the following form

\begin{equation}
H=
\begin{pmatrix}
A & \vline & JCJ \\
\hline 
C & \vline & JAJ 
\end{pmatrix} \ .
\end{equation}

\subsection*{Eigenvalues and Eigenvectors}

Centrosymmetric matrices have the property that their eigenvalues can be obtained from the diagonalization of specific combinations of the submatrices $A$ and $C$.  Further, they have either symmetric or antisymmetric eigenvectors. Consider an eigenvector $\textbf{v}$ corresponding to the eigenvalue $\lambda$ of the centrosymmetric matrix $H$

\begin{eqnarray}
H\textbf{v} &=& \lambda\textbf{v} \quad \quad | \cdot  J \rightarrow \nonumber
\\
JH\textbf{v} &=& \lambda J\textbf{v} \nonumber
\\
HJ\textbf{v} &=& \lambda J\textbf{v} \quad ,
\end{eqnarray}
where we used Eq.~\eqref{equ:def} and $J^2=\mathbb{1}$, it follows that $J\textbf{v}$ is also an eigenvector of H corresponding to the eigenvalue $\lambda$. Thus

\begin{eqnarray}
J\textbf{v} = a\textbf{v} \quad ,
\end{eqnarray}
holds, with $a \neq 0$ being the eigenvalue of J and since J is an orthogonal matrix, $a=\pm 1$. Hence, $\textbf{v}$ is either antisymmetric or symmetric. In our terms this means that

\begin{equation}
\textbf{v} = 
\begin{pmatrix}
v \\
\hline 
Jv 
\end{pmatrix}
\quad or \quad 
\begin{pmatrix}
v \\
\hline 
-Jv 
\end{pmatrix} 
\quad with \quad
\begin{pmatrix}
\text{\small{neg. Matsubara}} \\
\text{\small{frequencies}} \\
\hline 
\text{\small{pos. Matsubara}} \\
\text{\small{frequencies}}
\end{pmatrix}  
\quad ,
\end{equation}
where $\textbf{v}$ is a $2n\times 1$ vector and $v$ is a $n\times 1$ subpart of it.

Next we consider $\lambda_S$, an eigenvalue corresponding to a symmetric eigenvector $H \textbf{v}_S = \lambda_S \textbf{v}_S$

\begin{eqnarray}
\begin{pmatrix}
A & JCJ \\
C & JAJ 
\end{pmatrix}
\begin{pmatrix}
v \\
Jv 
\end{pmatrix}
&=& 
\lambda_S
\begin{pmatrix}
v \\
Jv 
\end{pmatrix}
\\
&\Downarrow& \nonumber
\\
(A+JC)v &=& \lambda_S v \ .
\end{eqnarray}
In a similar fashion one finds for $\lambda_A$, corresponding to an antisymmetric eigenvector

\begin{eqnarray}
\begin{pmatrix}
A & JCJ \\
C & JAJ 
\end{pmatrix}
\begin{pmatrix}
v \\
-Jv 
\end{pmatrix}
&=& 
\lambda_A
\begin{pmatrix}
v \\
-Jv 
\end{pmatrix}
\\
&\Downarrow& \nonumber
\\
(A-JC)v &=& \lambda_A v
\end{eqnarray}

This shows that the centrosymmetric matrix $H$ has eigenvalues $\lambda_S$ obtained from diagonalizing $A+JC$, which also gives the non-trivial parts $v$ of the symmetric eigenvectors $\textbf{v}_S$.
In our case they correspond to the orange and green divergence lines, for $\lambda_S=0$ (see main text). 
On the other hand we observe that $\lambda_A$ corresponds to antisymmetric eigenvectors obtained from the diagonalization of the submatrices $A-JC$.

In the following an elegant way to see this block structure of $H$ is presented, which will be used later in the proof.

\subsection*{Block-diagonalization}
\label{Appendix:BlockDiag}

Using the following orthogonal matrix $Q$ ($QQ^T=\mathbb{1}$)

\begin{equation}
Q=\frac{1}{\sqrt{2}}
\begin{pmatrix}
\mathbb{1}  & -J \\
\mathbb{1} & J 
\end{pmatrix}
\end{equation}
one can block-diagonalize a centrosymmetric matrix $H$
\begin{eqnarray}
QHQ^{T}&=&
\frac{1}{2}
\begin{pmatrix} \nonumber
\mathbb{1}  & -J \\
\mathbb{1} & J 
\end{pmatrix} 
\begin{pmatrix}
A & JCJ\\
C & JAJ 
\end{pmatrix}
\begin{pmatrix} 
\mathbb{1} & \mathbb{1}\\
-J & J 
\end{pmatrix}
\\
&=& 
\frac{1}{2}
\begin{pmatrix} 
\mathbb{1}  & -J \\
\mathbb{1} & J 
\end{pmatrix} 
\begin{pmatrix} \nonumber
A -JC & A + JC\\
C-JA & C+JA 
\end{pmatrix} \nonumber
\\
&=&  
\frac{1}{2}
\begin{pmatrix} \nonumber
2(A-JC) & \mathbb{0} \\
\mathbb{0} & 2(A +JC)
\end{pmatrix}
\\
&=&  
\begin{pmatrix}
A -JC & \mathbb{0} \\
\mathbb{0} & A +JC
\end{pmatrix} 
\label{equ:block}
\end{eqnarray}
revealing the previously discussed block  structure.

\subsection*{Bisymmetric Matrices}

As stated in the main text, due to the SU(2)- and the time-reversal-symmetry the centrosymmetric matrix $H$ considered is in fact bisymmetric. This has important consequences for the submatrices $A$ and $C$ introduced earlier

\begin{eqnarray}
H&=&H^T
\\
\begin{pmatrix}
A & JCJ\\
C & JAJ 
\end{pmatrix}
&=&
\begin{pmatrix}
A^T & C^T\\
(JCJ)^T & (JAJ)^T 
\end{pmatrix} \quad ,
\end{eqnarray}
as $J=J^T$ one finds $A=A^T$ immediately. For $C$ the following equation holds

\begin{equation}
\label{eq:skewC}
C^T = JCJ \rightarrow C^TJ^T = JC \rightarrow (JC)^T = JC \ .
\end{equation}

This means that the combination of submatrices yielding the eigenvalues and the corresponding 
symmetric or antisymmetric eigenvectors,
is symmetric, ensuring, together with the particle-hole symmetry, that the obtained eigenvalues are real

\begin{equation}
(A\pm JC)^T = A^T \pm (JC)^T \overset{\ref{eq:skewC}}{=} A \pm JC \quad \ .
\end{equation}

\section{The mapping of divergence lines}
\label{App:equality}

Because of the specific mapping from $U>0$ to 
 $U<0$ of $\chi_{\uparrow \uparrow}$ and $\chi_{\uparrow \downarrow}$, it is possible 
to show, that the red divergence lines for $U<0$ are the mirrored ones of $U>0$. 
It also follows that the symmetric density divergences $(U>0)$ are mapped to symmetric divergences in the magnetic channel for $U<0$. 

The starting point is to consider the bisymmetric $\chi_{\uparrow \uparrow}$ and $\chi_{\uparrow \downarrow}$ matrices, where the fermionic Matsubara frequency indices will be omitted in the 
following. 
$\chi_{\uparrow \uparrow}$ and $\chi_{\uparrow \downarrow}$ fulfill the following relations, discussed in the main text in 
Sec.~\ref{subsec:symmetries}, 
when mapped from positive to negative $U$

\begin{eqnarray}
\chi^{U>0}_{\uparrow \uparrow} &=& \chi^{U<0}_{\uparrow \uparrow} = \chi_{\uparrow \uparrow} =
\begin{pmatrix}
A & JBJ \\
B & JAJ
\end{pmatrix}
\\
\chi^{U>0}_{\uparrow \downarrow} &=& 
\begin{pmatrix}
C & JDJ \\
D & JCJ
\end{pmatrix} \quad \\
\chi^{U<0}_{\uparrow \downarrow} &=&
\chi^{U>0}_{\uparrow \downarrow} (-J) =
\begin{pmatrix}
C & JDJ \\
D & JCJ
\end{pmatrix}
\begin{pmatrix}
\mathbb{0} & -J \\
-J & \mathbb{0}
\end{pmatrix}
\nonumber \\
&=&
\begin{pmatrix}
-JD & -CJ \\
-JC & -DJ
\end{pmatrix} \ .
\end{eqnarray}

Block-diagonalization of $\chi_{\uparrow \uparrow}$ and $\chi_{\uparrow \downarrow}$ for both cases
leads to

\begin{eqnarray}
Q\chi_{\uparrow \uparrow}Q^T&=&
\begin{pmatrix}
A-JB & \mathbb{0} \\
\mathbb{0} & A+JB
\end{pmatrix}\\
Q\chi^{U>0}_{\uparrow \downarrow}Q^T&=&
\begin{pmatrix}
C-JD & \mathbb{0} \\
\mathbb{0} & C+JD
\end{pmatrix}\\
Q\chi^{U<0}_{\uparrow \downarrow}Q^T&=&
\begin{pmatrix}
-JD-J(-JC) & \mathbb{0} \\
\mathbb{0} & -JD+J(-JC)
\end{pmatrix}\nonumber \\ &=& 
\begin{pmatrix}
C-JD & \mathbb{0} \\
\mathbb{0} & -[C+JD]
\end{pmatrix} 
\end{eqnarray}

This shows immediately that the antisymmetric block of $\chi_{\uparrow\downarrow}$, ($C-JD$), is unchanged, whereas the symmetric one changes sign for $U>0 \leftrightarrow U<0$. 
Considering $\chi_d$ and $\chi_m$ for $U<0$ and $U>0$ and the relation
\begin{eqnarray}
Q\chi_{d^+,m^-}Q^T &=& Q(\chi_{\uparrow\uparrow} \pm \chi_{\uparrow\downarrow})Q^T \nonumber \\
&=& Q\chi_{\uparrow\uparrow}Q^T \pm Q\chi_{\uparrow\downarrow}Q^T \ ,
\end{eqnarray}
the following conclusions can be drawn:
\begin{eqnarray}
\label{eq:mapping_dens}
& & Q\chi_{d}^{U\gtrless 0}Q^{T}= \nonumber \\
& &
\begin{pmatrix}
[A\!-\!JB]\!+\![C\!-\!JD] & \mathbb{0} \\
\mathbb{0} & [A\!+\!JB]\!\pm\![C\!+\!JD]
\end{pmatrix}
\end{eqnarray}

\begin{eqnarray}
\label{eq:mapping_mag}
& & Q\chi_{m}^{U\gtrless 0}Q^{T} = \nonumber \\
& &
\begin{pmatrix}
[A\!-\!JB]\!-\![C\!-\!JD] & \mathbb{0} \\
\mathbb{0} & [A\!+\!JB]\!\mp\![C\!+\!JD]
\end{pmatrix} \ ,
\label{equ:neg_U}
\end{eqnarray}
where in the density case the $+$ sign corresponds to $U>0$ and the $-$ to $U<0$, while for the magnetic case the opposite order applies. From Eqs.\,(\ref{eq:mapping_dens},\ref{eq:mapping_mag}) the following three insights can be obtained:

$(i)$ The antisymmetric block of $Q\chi_{d}^{U\gtrless 0}Q^{T}$ is independent of the sign of $U$.
The diagonalization of $[A-JB] + [C-JD]$ yields the eigenvalues and the corresponding antisymmetric eigenvectors 
of $\chi_d$. Their singularity corresponds to a red divergence line, independent of the sign of $U$. This is the 
mathematical reason for the perfect mapping of the 
red divergence lines reported in Fig.~\ref{Fig:1} and the equality of the singular eigenvectors shown in Fig.~\ref{Fig:2} of the main text. Note that this statement is crucially dependent on the particle-hole 
symmetry of the problem. Otherwise the bisymmetry property is lost. 

$(ii)$ The antisymmetric block of $Q\chi_{m}^{U\gtrless 0}Q^{T}$ is also independent of the sign of $U$. This means that, irrespective of the sign of $U$, the eigenvalues corresponding to antisymmetric eigenvectors of $\chi_m$ can be calculated by diagonalizing $[A-JB] - [C-JD]$. However, so far none of these eigenvalues were found to be singular. 

$(iii)$ The symmetric parts of $\chi_d$ and $\chi_m$ are mapped in the following way:
$[A+JB]+[C+JD]$ is the symmetric blockmatrix of $\chi^{U>0}_d$ {\sl and} $\chi^{U<0}_m$. This explains why the 
symmetric density channel divergences for $U>0$ are mapped to divergences with symmetric eigenvectors in the 
magnetic channel for $U<0$. Analogously, $[A+JB]-[C+JD]$ is the symmetric blockmatrix of the enhanced channels $\chi^{U<0}_d$ {\sl and} 
$\chi^{U>0}_m$. Here the bisymmetry explains the mapping 
of the eigenvalues, as discussed in Sec.~\ref{sec:spectral}. 

Finally we note that also the matrix causing the divergences in the particle-particle up-down channel 
$\chi^{\nu (-\nu')}_{pp,\uparrow\downarrow} - \chi_{0,pp}^{\nu \nu'} $, is bisymmetric, having hence the same properties as mentioned above. Combining this insight with Eq.~\ref{eq:mapping_chi_pp} we have now fully clarified how the mapping of the generalized susceptibilities works and its exact relation with the physical degrees of freedom. 
The antisymmetric sectors are not mapped along the lines 
of Eq.~\ref{Eq:SpinMap}, but they cancel in the sum in Eq.~\ref{eq:ChiSpectrum}. The symmetric subparts on the other hand 
follow the mapping of the physical D.o.F..

\section{Susceptibility density in the binary mixture disordered model}
\label{App:BM}
We discuss here the susceptibility density $\rho_d(\chi)$, introduced in Sec.~\ref{sec:spectral}, for the Binary Mixture (BM) disordered case. This model is defined by the following Hamiltonian
\begin{equation}
H=-t\sum\limits_{<ij>}{c^{\dagger}_{i}c_{j}} + \sum_{i} \epsilon_i {c^{\dagger}_{i}c_{i}} \ . \label{eqn:BM}
\end{equation}
Here spin indices can be omitted and we can safely consider spinless electrons moving in a random background
with equal probability for $\epsilon_i=\pm W/2$.
\begin{figure}[b]
    \centering
    \includegraphics[width=0.47\textwidth]{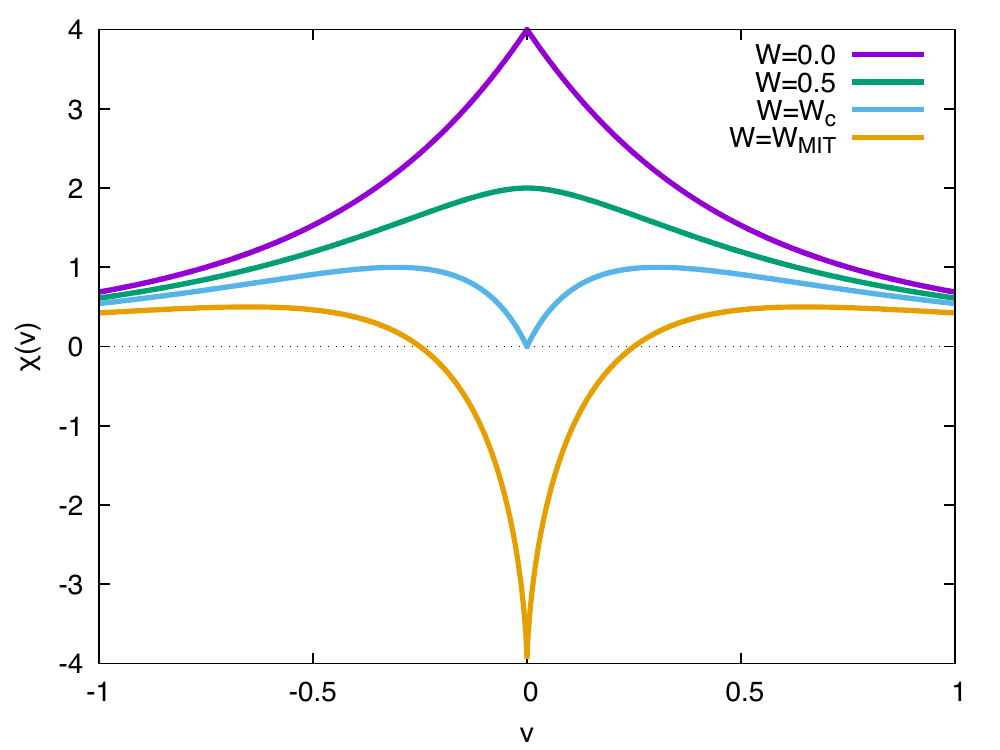}
    \caption{Eigenvalues of the generalized susceptibility of the DMFT solution of the BM model, from Eq.~(\ref{eq:chi}), with $D=1$.}
    \label{fig:chivsmatsu}
\end{figure}

The BM model can be readily solved in DMFT\cite{Freericks2003} where the corresponding expression for the Green's function at half-filling reads 
\begin{equation}
G(\nu)= \frac{1}{2} \left( \frac{1}{G^{-1}_0(\nu)-\frac{W}{2}}+\frac{1}{G^{-1}_0(\nu)+\frac{W}{2}}\right) \ ,
\label{eq:BM}
\end{equation}
with $G^{-1}_0(\nu)=\nu-D^2 G(\nu) / 4$ in the Bethe lattice case. This result is in perfect analogy with the Hubbard III (CPA)\cite{Soven1967} approximation for the Hubbard model, where $W$ must be understood as $U$.

Despite the simplicity of the model,  divergences of the irreducible vertex function as well as negative eigenvalues in the generalized susceptibility for the density channel are found also in the BM case\cite{Schaefer2013,Janis2014,Schaefer2016c}.  Specifically, the vertex divergence lines accumulate\cite{Schaefer2016c} at $T=0$ and $W_c/D=1/\sqrt{2}$, located well before the Mott-like transition ($W/D=1$).

The generalized susceptibility $\chi^{\nu\nu^\prime}_d$ of the BM model can be easily calculated at the DMFT level
\begin{equation}
\chi_{d}^{\nu \nu'}=\frac{ - 2 \beta }{\phantom{a} W^2}\sqrt{1+W^2G^2(\nu)}\left [ \sqrt{1+W^2G^2(\nu^\prime)}\mp 1\right] \delta_{\nu \nu'}
\label{eq:chi}
\end{equation}
where the $\pm$ sign is a consequence of the multivaluedness of the electronic self-energy and must hence be taken into account properly, in order to access the physical solution\cite{Schaefer2016c}.  We emphasize that Eq.~(\ref{eq:chi}) states that $\chi^{\nu\nu^\prime}_d$ is {\sl diagonal} in Matsubara frequency space, allowing for an immediate determination of the corresponding eigenvalues,  once the self-consistency condition has been enforced.
This is a consequence of the locality of the functional relation between the self-energy and the local single-particle propagator $\Sigma[G]$\cite{Schaefer2016c}.

According to Eq.~\eqref{eq:chi}, a singular eigenvalue  will occur when
\begin{equation}
1+W^2G^2=0
\label{eq:FKdivergence}
\end{equation}
for a given frequency $\nu$. The first vertex divergence is encountered when $1+W^2G^2=0$ for $\nu=0$.

The behavior of $\chi^{\nu=\nu^\prime}_d \equiv \chi(\nu)$ 
for different values of the disorder strength $W$ is shown in Fig.~\ref{fig:chivsmatsu}. 
Analytical estimates of the susceptibility density $\rho_d(\chi)$ can obtained from the analysis of the large and small frequency behavior of $\chi(\nu)$. 
In particular, in  the high frequency regime $\chi_d \propto   1/\nu^2$ holds. This is associated  to increasing number of  progressively smaller eigenvalues yielding an accumulation of positive contributions to the physical susceptibility around $\chi =0$. 
A similar accumulation can be found also in our DMFT calculations of  the Hubbard model, as it can be seen in Fig.~4 (Sec.~IV of the main text). The accumulation of high-frequency eigenvalues around $\chi \simeq 0$ is reflected in a Van Hove singularity of $\rho_d(\chi)$. In particular, by transforming the discrete summations into integrals in the large frequency domain, one finds that $\rho_d(\chi)\simeq \chi^{-3/2}$ around $\chi=0$.

At the same time, as one can see in Fig.~\ref{fig:chivsmatsu}, for small $W$ and small frequency $\chi(\nu)$ has a maximum. 
In the non interacting case, this is controlled by the scale $4/D^2$. 
Thus, in the weak disorder case, $\chi(\nu)$ has a compact support ranging from $\chi=0$ to a 
maximum value $\chi_{\rm Max} \propto 4/D^2 $. This is associated to a second Van Hove singularity, which
 changes to
$\rho_d(\chi)\simeq (\chi_{\rm Max}-\chi)^{-1/2}$ when it shifts to small but finite frequencies for $W/D \gtrsim 0.5$. 
This specific feature is due to the low-frequency behavior around the maximum value $\chi_{\rm Max}$.

\begin{figure}[t!]
    \centering
    \includegraphics[width=0.5\textwidth]{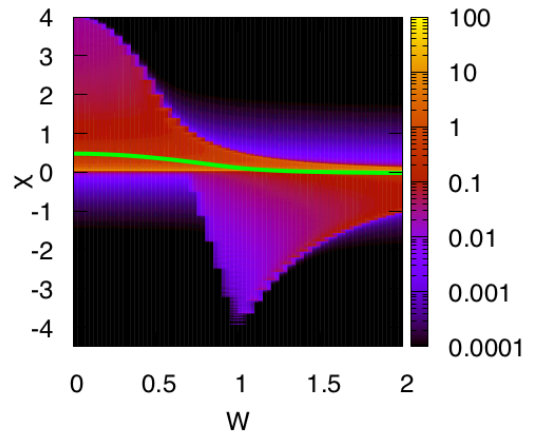}
    \caption{Intensity map (in logarithmic scale) of the susceptibility density $\rho_d(\chi)$ for the charge/density sector of the DMFT solution of the BM  model at $T=0$ as function of W.  The green curve displays the corresponding evolution of physical local susceptibility.}
    \label{fig:MapChiDistrib}
\end{figure}

In general, $\rho_d(\chi)$  changes considerably by increasing $W$ in the BM model, becoming finite also for {\sl negative} values of $\chi$, when $W>W_c = 1/\sqrt{2}$ ($D=1$). This happens before the Mott-like transition ($W=1$) occurs, where charge density fluctuation are frozen.

The overall behavior of $\rho_d(\chi)$ for the BM model at $T=0$, as well its relation with the corresponding evolution of the physical susceptibility,
 is summarized in Fig.~\ref{fig:MapChiDistrib}.
The values of $\rho_d(\chi)$, which becomes a continuous distribution in the $T=0$ limit, are plotted in a logarithmic intensity-color scale in Fig.~\ref{fig:MapChiDistrib}, which is necessary for highlighting the small weight associated to negative eigenvalues. 

The evolution of the physical susceptibility (cf.~Sec.~\ref{sec:spectral})
\begin{equation}
\langle \chi_d \rangle= \int d\chi  \,  \, \chi \, \rho_d(\chi),
\end{equation}
 is shown with a green continuous curve in Fig.~\ref{fig:MapChiDistrib} as a function of disorder strength $W$. 
 We see a crossover in this quantity, which progressively approaches zero with increasing $W$. 

At the two-particle level, the mechanism behind the gradual suppression of $\langle \chi_d \rangle$ looks different in the two physical phases (metallic for  $W < 1$ and insulating for $W >1$). In the former, one observes a progressive shift of the support $\chi_{\rm Max} -\chi_{\rm min}$ of $\rho_d(\chi)$ towards lower values of $\chi$, with $\chi_{\rm min}$ becoming negative for $W > W_c$, while the associated intensity does not change dramatically.
For $W\!>\!1$ the overall domain of  $\rho_d(\chi)$ progressively shrinks, because  $\chi_{\rm min} < 0 $ reaches its lowest  value at $W=1$ 
raising again for larger $W$, while $\chi_{\rm Max}> 0$ always decreases. As marked by the corresponding color change clearly noticeable in the right-half of the figure, the suppression of $\langle \chi_d \rangle$ for $W>1$ is mostly driven by an enhancement of $\rho_d(\chi)$ for negative $\chi$ values.
At finite temperature, in the case of a discrete spectrum, this corresponds to a progressive increase of the {\sl number} of negative eigenvalues of the generalizes susceptibility and to a simultaneous decrease of their absolute values, after the MIT is reached.

\bibliography{VERTEX}

\end{document}